\documentclass[11pt]{article}
\pdfoutput=1
\usepackage{jheppub}

\usepackage{blindtext}

\usepackage{mathrsfs}
\usepackage{bm}
\usepackage{paralist}
\usepackage{url}
\usepackage[]{microtype}

\usepackage{tikz-cd}
\usepackage{tikz}
\usepackage{pict2e}
\usepackage{float}
\usepackage{stmaryrd}
\usepackage{geometry}
\usepackage{tensor}
\usepackage{mathtools}

\usepackage{amsmath,amssymb,amsthm,amscd,graphicx}
\usepackage{psfrag}

\usepackage[utf8]{inputenc}
\usepackage{graphicx}
\usepackage{xcolor}
\usepackage{hyperref}
\hypersetup{colorlinks,allcolors=blue}

\definecolor{frangreen}{rgb}{0.040, 0.475, 0.435}

\usepackage[T1]{fontenc}

\usepackage[utf8]{inputenc}
\usepackage{graphicx}
\usepackage{amsmath,amsthm,amssymb,physics,mathtools}
\usepackage{tikz,tikz-cd, dynkin-diagrams}
\usepackage{epigraph}
\usepackage{hhline}
\usepackage{caption}
\usepackage{subcaption}
\usepackage{ytableau}
\usepackage{natbib}
\usetikzlibrary{positioning,arrows}
\usetikzlibrary{decorations.pathmorphing}
\usetikzlibrary{decorations.markings}
\usetikzlibrary{matrix}
\usepackage{bbold}

\usepackage{environ}         
\usepackage{etoolbox}        
\usepackage{graphicx}        
\usepackage{nomencl}
\makenomenclature

 \usepackage{etoolbox}
\renewcommand\nomgroup[1]{%
  \item[\bfseries
  \ifstrequal{#1}{C}{CFT symbols}{%
  \ifstrequal{#1}{H}{Heun symbols}{%
  \ifstrequal{#1}{F}{CFT symbols - semiclassics}{}}}%
]}
	\newlength{\myl}
\let\origequation=\equation
\let\origendequation=\endequation

\RenewEnviron{equation}{
  \settowidth{\myl}{$\BODY$}                       
  \origequation
  \ifdimcomp{\the\linewidth}{>}{\the\myl}
  {\ensuremath{\BODY}}                             
  {\resizebox{\linewidth}{!}{\ensuremath{\BODY}}}  
  \origendequation
}

\newcommand{\FIV}[9]{\mathfrak{F}\bigg( \begin{matrix} #2 \\ #1 \end{matrix} \, #3 \, \begin{matrix} #4  \\ \, \end{matrix} \,#5 \, \begin{matrix} #6 \\ #7 \end{matrix} ; #8, #9 \bigg)}

\newcommand{\FIVsc}[9]{\mathcal{F}\bigg( \begin{matrix} #2 \\ #1 \end{matrix} \, #3 \, \begin{matrix} #4  \\ \, \end{matrix} \,#5 \, \begin{matrix} #6 \\ #7 \end{matrix} ; #8, #9 \bigg)}

\newtheorem{remark}{Remark}[section]
\theoremstyle{definition}

\newcommand{\bea}{\begin{eqnarray}}
\newcommand{\eea}{\end{eqnarray}}

\def\v{\v}

\def\12{\frac{1}{2}}
\newcommand{\be}{\begin{equation}}
\newcommand{\ee}{\end{equation}}
\newcommand{\ba}{\begin{aligned}}
	\newcommand{\ben}{\begin{eqnarray}\displaystyle}
	\newcommand{\een}{\end{eqnarray}}

\makeatletter
\gdef\@fpheader{}
\makeatother
	
\begin{document}

		\title{One loop effective actions in Kerr-(A)dS Black Holes}
	
		\author{Paolo Arnaudo, Giulio Bonelli, Alessandro Tanzini}
		\affiliation{International School of Advanced Studies (SISSA), via Bonomea 265, 34136 Trieste, Italy}
		\affiliation{INFN, Sezione di Trieste, Trieste, Italy}
		\affiliation{Institute for Geometry and Physics, IGAP, via Beirut 2, 34136 Trieste, Italy}
    \emailAdd{parnaudo@sissa.it}
    \emailAdd{bonelli@sissa.it}
    \emailAdd{tanzini@sissa.it}
		
		\date{\today}

\abstract{
We compute new exact analytic expressions for one-loop scalar effective actions in Kerr (A)dS black hole (BH) backgrounds in four and five dimensions.
These are computed by the connection coefficients of the Heun equation via a generalization of the Gelfand-Yaglom formalism to second-order linear ODEs with regular singularities.
The expressions we find are in terms of Nekrasov-Shatashvili special functions, 
making explicit the analytic properties of the one-loop effective actions with respect to the gravitational parameters and the precise contributions of the quasi-normal modes. 
The latter arise via an associated integrable system.
In particular, 
we prove asymptotic formulae for large angular momenta in terms of hypergeometric functions and
give a precise mathematical meaning to Rindler-like region contributions. 
Moreover, we identify the leading terms in the large distance expansion as the point particle approximation of the BH and their finite size corrections as encoding the BH tidal response.
We also discuss the exact properties of the thermal version of the BH effective actions by providing a proof of the DHS formula and explicitly computing it for new relevant cases.
Although we focus on the real scalar field in dS-Kerr and (A)dS-Schwarzschild in four and five dimensions, similar formulae can be given for higher spin matter and radiation fields in more general gravitational backgrounds.
}

\maketitle

\section{Introduction}

The study of quantum field theory in curved spaces has many important applications ranging from more formal ones such as AdS/CFT correspondence to more phenomenological such as the study of the production (and the propagation) of gravitational waves by black holes (BH) -- and more in general heavy compact objects -- as well as the behavior of radiation and quantum matter in their vicinity, determining for example the properties of their photo-sphere. In order to obtain {\it exact analytic results} in this framework, the first approach is to start from the analysis of BH linear perturbation theory while postponing that of non-linear effects. These might be important to understand fine structure effects in gravity and systems coupled to it.

The linearization of Einstein equations around BH solutions has been a classical subject of investigation in parallel with that of the study of one-loop effective actions in BH metric backgrounds.
As it is well known, the high degree of symmetry of many BH solutions allows the reduction of the problem to second-order ODEs, namely the Teukolsky equation
\cite{teukolsky1972}, due to the separation of variables. The relevant differential operators are second-order linear operators related to Fuchsian 
equations with four regular singularities on the Riemann sphere, also known as the Heun equation. Notice that this holds for Petrov type D metrics in general \cite{Batic:2007it}.

Recently an exact analytic formalism to handle spinning (A)dS BH has been developed \cite{Aminov:2020yma,Bonelli:2021uvf,Bianchi:2021mft,Bianchi:2021xpr} from a powerful technique arising in the context of supersymmetric quantum field theory in four dimensions and conformal field theory in two dimensions.
This is based on the Nekrasov instanton counting formulas in $D=4$ ${\cal N}=2$ gauge theories \cite{Nekrasov:2002qd}, the Alday-Gaiotto-Tachikawa correspondence
with Liouville conformal field theory (CFT) \cite{Alday:2009fs}, and the link of these to integrable systems developed by Nekrasov and Shatashvili \cite{ns}.
In this context, the semiclassical limit of Liouville correlation functions concretely realizes the explicit solution of the Heun differential equation, and crossing-symmetry of 2d CFT allows to compute the matrix of connection coefficients in terms of the so-called Nekrasov-Shatashvili (NS) function \cite{Bonelli:2022ten}. See also \cite{daCunha:2021jkm} for parallel speculations.

In this paper, we apply the above techniques to compute the one-loop Euclidean quantum action of scalar particles in full four and five-dimensional Kerr-(A)dS backgrounds, whose 
field equations separate in Heun equations. Let us observe, moreover, that the same techniques can be applied in higher dimensions, which typically give rise to ODEs with more regular and/or irregular singularities.

A general formalism for the study of the quantum effective actions in a black hole background in Euclidean quantum gravity \cite{PhysRevD.15.2752} was settled in \cite{Denef:2009kn}, where a formula for the computation of determinants in thermal spacetimes in terms of quasinormal modes was proposed\footnote{A related observation was done in \cite{Dodelson:2023vrw}
where the propagator in Lorentzian signature was proposed
to have a similar structure, the one-loop effective action being the integral of the logarithm of the Fourier transformed propagator.}. However, explicit calculations were so far restricted to three-dimensional BHs or more in general problems reducible to Hypergeometric operators. The general structure of the determinant was conjectured to be given as a product over the (anti)quasinormal modes of the BH background. We provide an explicit calculation and proof of this statement in this paper, by making use of an improved computational technique by which the relevant analytic properties of the 
effective actions are manifest.
We devise a method to compute the one-loop determinants directly from the connection coefficients of the Heun equations
by generalizing the Gelfand-Yaglom method \cite{Gelfand:1959nq}
to differential operators with singularities. These results are obtained via simple manipulations in Appendix~\ref{appendixGY} and then applied in the paper to the Heun differential 
operator. Similar formulas appear also in \cite{lesch1999determinants}.
The application of the Gelfand-Yaglom method to higher dimensional differential operators
admitting separation of variables is not a novelty, as it has been studied in 
the elegant analysis of \cite{Dunne:2007rt}.
 
Our computations make contact with previous results as follows. 
As it is well known \cite{Denef:2009kn, Law:2022zdq}, the quantum effective actions are parameterized in terms 
of the QNMs up to the exponent of a polynomial function in the background parameters
(see also \cite{Grewal:2022hlo} for more recent developments).
In this paper, we give closed formulas for the determinants leading to the one-loop effective action and
the rule to compute their spectrum from the NS function of the quantum integrable system associated to the specific Heun equation arising in the gravitational model
in the above perspective.

An interesting feature of our results is that the formulae in the full BH background, when computed at the leading order in perturbation theory in the BH radius and for large enough angular momenta of the perturbation, undergo a simplification corresponding to the point-like approximation for the BH in the sense of \cite{Bautista:2023sdf}. In this approximation, the one-loop effective action can be obtained by approximating to a Hypergeometric problem with shifted parameters, which explicitly change the arguments of the Gamma functions
(see formulae \eqref{1} and \eqref{2}). The complete determinant is instead computed by taking into account the full extended 
structure of the horizon which is relevant to keep track of the tidal properties of the BH and is given in \eqref{3}.

Moreover, the formula for the determinant of the Sturm-Liouville operator with two regular singularities at $z=0$ and $z=1$ has a clear physical meaning when specialized to the near-horizon analysis as in \cite{Law:2022zdq}. 
We indeed get 
\begin{equation}\label{zero}
\mathrm{det}\left(\frac{\mathrm{d}^2}{\mathrm{d}z^2}+V(z)\right)=
\frac{2\pi}{\Gamma\left(1+a_0+a_1\right)\Gamma\left(a_0+a_1\right)}
\frac{\mathcal{C}_{12}}{\widetilde{\mathcal{C}}_{12}}
\end{equation}
where $a_0$ and $a_1$ (supposed to have a positive real part) are the indices of the two singular points, $\mathcal{C}$ denotes the Heun connection matrix between two bases of independent normalized solutions around $z=0$ and $z=1$, $\mathcal{C}_{12}$ being the connection coefficient in front of the discarded local solution at $z=1$, and $\widetilde{\mathcal{C}}_{12}$ denotes the corresponding connection coefficient in a reference simplified Hypergeometric problem whose singularities at $z=0$ and $z=1$ have the same indices (see Appendix~\ref{appendixHeunoverhyperg} for the derivation of the above formula).
The factors in the denominator coincide with the ones that were used in \cite{Law:2022zdq} to normalize the effective action to extract the physical single-particle density of states of an ideal thermal gas of scalar particles in the BH background. This physically correspond to the normalization of the effective action due to the Rindler-like region, which is obtained by zeta-regularizing the exact hypergeometric computation. This gives a direct explanation and explicit proof of the observation made in \cite{Law:2022zdq} that the Euclidean path integral uniquely fixes the reference for normalization. 

By applying the results of \cite{Bonelli:2022ten} for the connection matrices\footnote{See also
\cite{Lisovyy:2022flm} for a mathematical counterpart of the analysis.}
in \eqref{zero} and the proper dictionary between gauge theory and gravity variables, one can use the 
above determinant formula to compute the one-loop effective action in Petrov type D BH (or their thermal counterpart) backgrounds.
Concretely, after the separation of variables, the 
determinants 
of the radial modes at given frequency and angular quantum numbers 
is given in terms of the Nekrasov-Shatashvili function $F$ as
\begin{equation}\label{one}
\begin{aligned}
\mathrm{det}(\mathcal{D}_{\mathrm{rad}}-A_{\ell \vec{m}})[\omega]=&\sum_{\theta'=\pm}\frac{2\pi\Gamma(-2\theta'a)\Gamma(1-2\theta'a)}{\prod_{\sigma=\pm}\Gamma\left(\frac{1}{2}- a_0-\theta'a+\sigma\,a_t\right)\Gamma\left(\frac{1}{2}-\theta'a+ a_1+\sigma\,a_{\infty}\right)}\times\\
&\times t^{- a_0+\theta' a}\exp(-\frac{1}{2}\partial_{a_0}F(t)+\frac{1}{2}\partial_{a_1}F(t)-\frac{\theta'}{2}\partial_aF(t)).
\end{aligned}
\end{equation}
These get re-summed in the quantum numbers (see \eqref{fulldet}) leaving behind the one-loop BH effective action 
\begin{equation}\label{two}
\begin{aligned}
\Gamma^{1-loop}_{\mathrm{BH}}=\int_{-\infty}^{\infty}&\mathrm{d}\omega\sum_{\ell,\vec{m}}{\log}
\left(
\mathrm{det}(\mathcal{D}_{\mathrm{rad}}-A_{\ell \vec{m}})[\omega]
\right).
\end{aligned}
\end{equation}
In the above formula, we used the fact that the contributions of the QNMs coincide with those of the anti-QNMs (see Sec.~\ref{conclusion} for a detailed explanation), and the factor $1/2$ coming from the functional determinant of the Klein-Gordon operator for the real scalar simplifies. 
The concrete dictionaries connecting the gauge theory-like variables in 
\eqref{one}
and the gravitational variables of the specific cases we study
are in \eqref{KerrdSdictionary2} for the Kerr-de Sitter case, in \eqref{SdSdictionary} for the Schwarzschild-de Sitter case, and in \eqref{SAdSdictionary} for the Schwarzschild-anti-de Sitter case. 
Note that the formula itself also depends on the dictionary through the sign of $\mathrm{Re}(a_0)$ and $\mathrm{Re}(a_1)$ (see Appendix~\eqref{determinantofhyperg}).
In equation \eqref{one}, $F(t)$ is a new special function which is explicitly calculable and is crucial for the solution of the Heun connection problem. We briefly review its definition and calculation in Appendix~\ref{appendixNS}. 
As anticipated, the above exact formulae reduce to simpler
Hypergeometric-like ones if one of the two channels labeled by $\theta'$ and the contribution of exponential factors can be neglected in the evaluation of the quasi-normal modes. 
This corresponds to the point-particle approximation for the black hole, while the neglected 
channel takes into account the finite-size effects of the black hole itself.

After Wick rotation to the thermal circle, the one-loop effective action can be computed via $\zeta$-function regularization from
\begin{equation}\label{zetabh}
\zeta_{\mathrm{BH}}(s) = \frac{1}{\Gamma(s)} \int_0^\infty\frac{dt}{t}\, t^s\, \frac{1+e^{-\beta t}}{1-e^{-\beta t}}
\sum_{\mathrm{z\in QNM}} d_{z} e^{-zt} 
\end{equation}
the regularised physical QNM character of the thermal space-time being given by 
$-\zeta_{\mathrm{BH}}'(0)$. 
The QNMs in \eqref{zetabh} get explicitly computed via the associated quantum integrable system
as explained below, with $d_z$ being the associated degeneracy.
In particular, 
the zeros of \eqref{KerrdSdeteta} provide the (anti)QNMs in the 4D Kerr-dS case
and 
those of \eqref{5dSads} those in the 5D Schwarzschild-AdS case.

The content of the paper is the following. In Sec.~\ref{section2} we discuss the technique we use to compute the relevant determinants via an extension of the Gelfand-Yaglom theory to second-order operators with regular singularities. In Sec.~\ref{section3} we apply the above method to the computation of the one-loop effective action in the Kerr-de Sitter background in four dimensions and we discuss its limits to Schwarzschild-de Sitter and pure de Sitter cases. In Sec.~\ref{section4} we consider the five-dimensional Schwarzschild-Anti de Sitter background
and discuss its limit to pure AdS$_5$. Finally, in Sec.~\ref{conclusion} we discuss the thermal version of our results via Wick rotation and compactification on the circle of \eqref{two} and we compare 
the leading expression in the BH radius expansion with previous results in the literature. In the Appendices we discuss more mathematical aspects of GY theorem and its applications and we provide a brief account of the gauge theory formulae for the NS function which are relevant to the computations in the main text. 

Let us close this introductory section by discussing few open points for further explorations.
In this paper, we obtained exact analytic formulae for the one-loop effective actions of a scalar field in BH backgrounds. We remark that these formulae display the explicit contribution of each quasi-normal mode in an analytic form and are therefore amenable to single out the contribution of individual ones to study their physical relevance.
It would be also very important to fully exploit the properties of the concrete expressions we get and their physical content in order to better understand the behaviour of quantum matter in presence of stationary BHs.
Similar analysis have been initiated in parallel for Kerr-Compton scattering amplitudes
in \cite{Bautista:2023sdf}. For later related developments see
\cite{Ivanov:2024sds,Cangemi:2023bpe}.

Let us also remark that the extension of the Gelfand-Yaglom theorem to operators with regular singularities presented in this paper opens the possibility to numerically evaluate the determinants from the properly normalized Heun functions, whose numerical values are available in Mathematica.
For example, the numerical implementation of Heun functions is described in \cite{Hatsuda:2020sbn}, where also QNMs of Kerr-de Sitter black holes are computed.

It would be interesting to investigate higher spin perturbations within this method\footnote{See \cite{Grewal:2022hlo} for a discussion about one-loop effective actions in this case.}.
Perturbations up to spin two of the four-dimensional Kerr black hole were studied in \cite{Bonelli:2022ten}. 
These results can be used to describe the one-loop Euclidean gravitational partition function and complement the analysis of 
logarithmic corrections to Kerr and Kerr-AdS BH thermodynamics recently studied in 
\cite{Kapec:2023ruw,Maulik:2024dwq}. For recent studies on quantum effects and their consequence on the thermodynamic behavior of near-extremal black holes see also \cite{Kapec:2024zdj,Kolanowski:2024zrq}.
We remark that one-loop effective actions in supergravity were used to compute logarithmic corrections to the entropy of supersymmetric BHs \cite{Banerjee:2010qc,Sen:2014aja}. By complementing the analysis performed in this paper with the study of the zero-modes of the one-loop operators one should be able to compute logarithmics corrections to the entropy of Kerr-(A)dS BHs.

Our computations are performed in zeta-function regularization, it would be interesting to compare with direct $\overline{\mathrm{MS}}$ Feynman diagram dimensional regularization procedure, as done in \cite{Dunne:2005rt}
in the context of false vacuum decay calculations.

Other gravitational backgrounds can be analyzed with the same methods, see
\cite{Bianchi:2021mft,Bianchi:2021xpr,Bianchi:2021yqs,Bianchi:2022qph, Bianchi:2022wku,Bianchi:2023lrg,Bianchi:2023rlt,Bianchi:2023sfs,Bianchi2021MoreOT,Amado:2021erf,Imaizumi:2022qbi,Giusto:2023awo,Cipriani:2024ygw} for QNMs analysis,
and it should therefore be possible to extend our method to the study of one-loop effective actions also in these cases (see also \cite{Zhao:2024rrw}).
Related parallel investigations on QNMs appear in
\cite{Bucciotti:2023ets,Lei:2023mqx,Grozdanov:2023tag,Hatsuda:2023geo,Hatsuda:2021gtn,Kehagias:2022ndy}.
As the results in the realm of micro-state counting have been encouraging, these should also apply to the framework considered in \cite{Krishnan:2023jqn}.
On the other hand, effective near-horizon symmetries 
have been analyzed in \cite{Charalambous:2022rre,Charalambous:2024tdj} and it could be interesting to consider them in the light of our results.

We remark that the same methods have been applied also to the study holographic thermal correlation functions for five-dimensional AdS BHs in the large N limit \cite{Dodelson:2022yvn,Dodelson:2023vrw,Bhatta:2023qcl,He:2023wcs,Berenstein:2022nlj} and for four-dimensional AdS BH
in the hydrodynamical limit \cite{Aminov:2023jve},
while the one-loop contributions computed here should allow to investigate the leading 1/N corrections. 

Let us remark that the instanton series used to compute perturbatively the Nekrasov function relevant for the problems studied in this paper have finite radius of convergence \cite{Arnaudo:2022ivo}. Although its Nekrasov-Shatashvili limit is believed to share the same convergence properties, it would be important to establish this rigorously via a direct analysis.
Let us also mention that other methods exist to evaluate the NS function in terms of Fredholm determinants and of TBA equations \cite{Bonelli:2016idi,
Grassi:2019coc, Fioravanti:2019vxi}, and they have been applied to the study of BH QNMs \cite{Fioravanti:2021dce,Fioravanti:2022bqf}.
It would be useful to use these methods also in the evaluation of BH effective actions.

A relevant question in this context is about the completeness of the quasinormal modes \cite{Ching:1995rt,Beyer:1998nu,Nollert:1996rf,Nollert:1998ys,Berti:2009kk,Jafferis:2013qia,London:2023aeo} and also about the pseudo-spectrum of these differential operators. The latter is relevant to discuss the stability of BHs against linear perturbations and the stability of the QNMs against perturbations of the potential of the wave equation (see for example \cite{Arean:2023ejh,Sarkar:2023rhp,Cownden:2023dam,Casals:2021ugr}).

As the original results of \cite{Bonelli:2021uvf}
were giving also exact information on the tidal Love number -- see also \cite{Pereniguez:2021xcj} -- of the BH, it would be interesting to consider
the analogue computational problem for the spinning de-Sitter BH case.

{\bf Acknowledgments:} 
We would like to thank 
Gleb Aminov, 
Massimo Bianchi, Gerald Dunne, Francesco Fucito,
Alba Grassi, Cristoforo Iossa, Francisco Morales, Zihan Zhou
for useful discussions and comments.

The research of G.B. is partly supported by the INFN Iniziativa Specifica ST\&FI and by the PRIN project “Non-perturbative Aspects Of Gauge Theories And Strings”. The research of  P.A. and A.T. is partly supported by the INFN Iniziativa Specifica GAST and InDAM GNFM. 
The research is partly supported by the MIUR PRIN Grant 2020KR4KN2 "String Theory as a bridge between Gauge Theories and Quantum Gravity".  
All the authors acknowledge funding from the EU project Caligola HORIZON-MSCA-2021-SE-01), Project ID: 101086123, and CA21109 - COST Action CaLISTA. 
G.B. thanks the Galileo Galilei Institute for Theoretical Physics for the hospitality and the INFN for partial support during the final stages of this work.

\vspace{1cm}

\section{One-loop black hole effective actions and Gelfand-Yaglom theorem}\label{section2}
 
In this section we apply the Gelfand-Yaglom theorem (see Appendix~\ref{appendixGY}) to compute determinants of the second-order separable differential operators which 
compute the one-loop effective actions in the BH backgrounds.
More precisely, we consider the conformally coupled Klein-Gordon differential operator in four-dimensional Kerr-de Sitter black holes and the Klein-Gordon differential operator in Schwarzschild anti-de Sitter black holes in five dimensions. In both cases, 
the action of the scalar field can be written as
\begin{equation}
S[\Phi,g_{\mu\nu}]=\int\,\mathrm{d}^Dx\,\sqrt{-g}\,\left(-\frac{1}{2}g^{\mu\nu}\nabla_{\mu}\Phi\,\nabla_{\nu}\Phi-\frac{1}{2}\mu^2\Phi^2\right),
\end{equation}
where $g_{\mu\nu}$ is the metric of the spacetime, and the resulting differential operator can be written as
\begin{equation}
\left[\frac{1}{\sqrt{-g}}\partial_{\mu}\left(\sqrt{-g}\,g^{\mu\nu}\partial_{\nu}\right)-\mu^2\right]\Phi\equiv\left[\Box-\mu^2\right]\Phi=0,
\end{equation}
where, for Kerr-de Sitter BH in four dimensions, we fix $\mu^2=2$, whereas, for Schwarzschild anti-de Sitter black hole in five dimensions, we consider $\mu$ to be generic. In the latter case, it is convenient to reparametrize $\mu$ as
\begin{equation}
\mu^2=\Delta(\Delta-4),
\end{equation}
where $\Delta$ corresponds to the conformal dimension of the scalar field living in the holographic dual 4$d$ CFT. We require $\Delta\notin\mathbb{Z}$ in order to avoid logarithmic solutions for the radial function around the AdS boundary. 

The Gelfand-Yaglom theorem provides a way to compute the logarithm of the inverse of the partition functions associated with the above Klein-Gordon differential operators.
The computation of the full determinant $\mathrm{det}\left(\Box-\mu^2\right)$ can be reduced to the computation of the determinant of a radial 1-dimensional operator which depends on the eigenvalues of the other separated problems and their degeneracies. We use the following decomposition in Fourier modes of the wave function $\Phi$ 
\begin{equation}\label{decomposition}
\Phi(t,r,\Omega)=\int_{-\infty}^{\infty}\mathrm{d}\omega\sum_{\ell,\vec{m}}e^{-i\omega t}S_{\omega,\ell,\vec{m}}(\Omega)R_{\omega,\ell,\vec{m}}(r).
\end{equation}
In the spherically symmetric cases, the angular functions $S_{\omega,\ell,\vec{m}}(\Omega)$ coincide with the spherical harmonics $Y_{\ell,\vec{m}}(\Omega)$.
Starting from the problem
\begin{equation}
\left(\Box-\mu^2\right)\Phi=\lambda\Phi,
\end{equation}
and using \eqref{decomposition}, we obtain a system of coupled second-order differential equations for $S_{\omega,\ell,\vec{m}}(\Omega)$ and $R_{\omega,\ell,\vec{m}}(r)$, of the form
\begin{equation}
\begin{aligned}
&\mathcal{D}_{\mathrm{rad}}R_{\omega,\ell,\vec{m}}(r)=\left(A_{\ell \vec{m}}+\lambda\right)R_{\omega,\ell,\vec{m}}(r),\\
&\mathcal{D}_{\mathrm{ang}}S_{\omega,\ell,\vec{m}}(\Omega)=-A_{\ell \vec{m}}\,S_{\omega,\ell,\vec{m}}(\Omega),
\end{aligned}
\end{equation}
for some second-order differential operators $\mathcal{D}_{\mathrm{rad}}$ and $\mathcal{D}_{\mathrm{ang}}$, and where $A_{\ell \vec{m}}$ denotes the separation constant at fixed values of the quantum numbers.

The expression of the separation constant is obtained from the angular equation and then, when substituted into the radial equation, gives the determinant in terms of $\omega$ and the quantum numbers. In the Kerr-de Sitter case, the separation constant is expressed as an instanton expansion in terms of NS functions. In the Schwarzschild-(anti-)de Sitter cases, the angular eigenfunctions reduce to the spherical harmonics and the separation constant has an exact expression in terms of the quantum number $\ell$.

In the asymptotically de Sitter black hole problems, around the points in which the boundary conditions are imposed  -- which are horizons of the BH geometry --  a basis of independent solutions of the radial equation behaves like
\begin{equation}
R_{\omega,\ell,\vec{m}}(r)\sim\exp(\pm i\omega r_*),
\end{equation}
where $r_*$ is the tortoise coordinate. When $\omega$ is analytically continued to assume values in the complex plane, the 
boundary conditions select the correct local solutions according to the sign of the imaginary part of $\omega$.
In the asymptotically anti-de Sitter black hole problem, this still holds for the boundary condition imposed around the black hole horizon, but the second boundary condition is imposed at the AdS boundary which is a regular point, and the selected solution depends on the value of the mass of the scalar perturbation.

The full determinant has an expression of the form
\begin{equation}\label{fulldet}
\log(\mathrm{det}\left(\Box-\mu^2\right))\equiv \int_{-\infty}^{\infty}\mathrm{d}\omega\sum_{\ell,\vec{m}}\log\left(\mathrm{det}\left(\mathcal{D}_{\mathrm{rad}}-A_{\ell \vec{m}}\right)[\omega]\right).
\end{equation}



When applying the Gelfand-Yaglom theorem to the 1-dimensional radial operator, we introduce a new variable $z$ such that the radial differential equation can be brought in Heun's form 
\begin{equation}\label{heunnormalform}
\quad\frac{\mathrm{d}^2\,\psi(z)}{\mathrm{d}\,z^2} + \left[\frac{\frac{1}{4}-a_0^2}{z^2}+\frac{\frac{1}{4}-a_1^2}{(z-1)^2} + \frac{\frac{1}{4}-a_t^2}{(z-t)^2}- \frac{\frac{1}{2}-a_1^2 -a_t^2 -a_0^2 +a_\infty^2 + u}{z(z-1)}+\frac{u}{z(z-t)} \right]\psi(z)=0,
\end{equation}
and $z=0$ and $z=1$ become the two \emph{singular} points in which we impose the boundary conditions. Let
\begin{equation}
\psi^{(\hat{z})}_{i,\lambda}(z)=(z-\hat{z})^{\frac{1}{2}\pm a_{\hat{z}}}\left[1+\mathcal{O}(z-\hat{z})\right],\quad i=1,2
\end{equation}
be the fundamental system of local solutions around $z=\hat{z}$.
The solution selected by the boundary condition at $z=\hat{z}$ is the one having in front of the exponent $a_{\hat{z}}$ the same sign of $\mathrm{Re}(a_{\hat{z}})$. For the problems we consider, this condition changes according to the values of the gravitational quantities. In particular, around the singularities corresponding to horizons of the geometry, the condition depends on the sign of $\mathrm{Im}(\omega)$, when this is analytically continued to take values on the complex plane.

Let us denote with $\psi^{(\hat{z})}_{1,\lambda}(z)$ the solution selected by the boundary condition at $z=\hat{z}$. 
Using the connection formulae, we can write
\begin{equation}
\psi^{(0)}_{1,\lambda}(z)=\mathcal{C}_{11,\lambda}\psi^{(1)}_{1,\lambda}(z)+\mathcal{C}_{12,\lambda}\psi^{(1)}_{2,\lambda}(z),
\end{equation}
where we denote with $\mathcal{C}_{11,\lambda},\mathcal{C}_{12,\lambda}$ the connection coefficients, which depend on $\lambda$ (but are independent of $z$).

In order to apply the Gelfand-Yaglom theorem, we introduce a reference problem whose differential operator $\tilde{\mathcal{D}}_{\mathrm{rad}}$ is a Hypergeometric one, obtained by simplifying the Heun differential equation keeping the indices of the singular points at $z=0$ and $z=1$ fixed. 
When computing ratios of the determinants of the two radial operators (the one of the original problem and the one of the reference problem), we have
\begin{equation}\label{radialproblem1}
\frac{\mathrm{det}\left(\mathcal{D}_{\mathrm{rad}}-A_{\ell m}-\lambda\right)}{\mathrm{det}\left(\tilde{\mathcal{D}}_{\mathrm{rad}}-\lambda\right)}\propto\frac{\mathcal{C}_{12,\lambda}}{\tilde{\mathcal{C}}_{12,\lambda}},
\end{equation}
where $\tilde{\mathcal{C}}$ denotes the connection matrix of the reference problem.
The above statement holds since both the LHS and the RHS (as functions of $\lambda$) have zeros in the eigenvalues of $\mathcal{D}_{\mathrm{rad}}-A_{\ell m}$ and poles in the eigenvalues of $\tilde{\mathcal{D}}_{\mathrm{rad}}$. 
The fact that the connection coefficient $\mathcal{C}_{12,\lambda}$ has zeroes in the eigenvalues is due to the fact that, if $\hat{\lambda}$ is an eigenvalue, then $\psi_{1,\lambda=\hat{\lambda}}^{(1)}(1)=0$ because of the boundary condition, and $\psi_{1,\hat{\lambda}}^{(0)}(1)=0$ if and only if $\mathcal{C}_{12,\hat{\lambda}}=0$.
Moreover, in the limit $\lambda\to\infty$ the ratio \eqref{radialproblem1} tends to 1. We thus 
conclude that
\begin{equation}\label{radialproblem2}
\frac{\mathrm{det}\left(\mathcal{D}_{\mathrm{rad}}-A_{\ell m}\right)}{\mathrm{det}\left(\tilde{\mathcal{D}}_{\mathrm{rad}}\right)}=\frac{\mathcal{C}_{12,\lambda=0}}{\tilde{\mathcal{C}}_{12,\lambda=0}}.
\end{equation}
Finally, as described in Appendix~\ref{appendixHeunoverhyperg}, we can compute the regularized determinant for the reference Hypergeometric potential. This provides a solution for the determinant of the radial Heun differential operator, which is of the form
\begin{equation}
\mathrm{det}\left(\mathcal{D}_{\mathrm{rad}}-A_{\ell m}\right)=2\pi\frac{\mathcal{C}_{12,\lambda=0}}{\Gamma(1+2\theta_0a_0)\Gamma(2\theta_1a_1)},
\end{equation}
as obtained in Appendix~\ref{appendixHeundeterminant}, where $a_0,a_1$ denote the indices of the singularities at $z=0$ and $z=1$ of the Heun differential operator, and where $\theta_0,\theta_1=\pm$, the signs being the same of the ones of the real parts of the indices $a_0,a_1$, respectively.

In the following sections, we concretely compute the determinants for the gravitational problems, and in the last section (see Section \ref{conclusion}) we discuss the results and rewrite the previous formulae more explicitly.

\section{Kerr-de Sitter spacetime in four dimensions}\label{section3}

The four-dimensional Kerr-de Sitter metric in Chambers-Moss coordinates can be written as
\begin{equation}\label{KerrdSmetric}
\begin{aligned}
 \mathrm{d}s^2&=\frac{r^2+x^2}{\Delta_r}\mathrm{d}r^2+\frac{r^2+x^2}{(a_{\mathrm{BH}}^2-x^2)\bigl(1+\frac{\Lambda}{3}x^2\bigr)}\mathrm{d}x^2+\\
&-\frac{\Delta_r-(a_{\mathrm{BH}}^2-x^2)\bigl(1+\frac{\Lambda}{3}x^2\bigr)}{(r^2+x^2)\bigl(1+\frac{\Lambda a_{\mathrm{BH}}^2}{3}\bigr)^2}\mathrm{d}t^2+\\
&+\frac{(a_{\mathrm{BH}}^2-x^2)}{a_{\mathrm{BH}}^2(r^2+x^2)\bigl(1+\frac{\Lambda}{3}a_{\mathrm{BH}}^2\bigr)^2}\biggl[(r^2+a_{\mathrm{BH}}^2)^2\biggl(1+\frac{\Lambda}{3}x^2\biggr)-(a_{\mathrm{BH}}^2-x^2)\Delta_r\biggr]\mathrm{d}\phi^2+\\
&+2\frac{(a_{\mathrm{BH}}^2-x^2)}{a_{\mathrm{BH}}(r^2+x^2)\bigl(1+\frac{\Lambda}{3}a_{\mathrm{BH}}^2\bigr)^2}\biggl[\Delta_r-(r^2+a_{\mathrm{BH}}^2)\biggl(1+\frac{\Lambda}{3}x^2\biggr)\biggr]\mathrm{d}t\,\mathrm{d}\phi,
\end{aligned}
\end{equation}
where
\begin{equation}
\begin{aligned}
x&=a_{\mathrm{BH}}\cos\theta,\\
\Delta_r(r)&=r^2-2Mr+a_{\mathrm{BH}}^2-\frac{\Lambda}{3}r^2(r^2+a_{\mathrm{BH}}^2)=-\frac{\Lambda}{3}(r-R_+)(r-R_-)(r-R_h)(r-R_i).
\end{aligned}
\end{equation}
In the previous equations, $M$ is the mass parameter of the black hole, $a_{\mathrm{BH}}$ is the parameter characterizing its angular momentum, $\Lambda>0$ is the cosmological constant, and we have factorized $\Delta_r(r)$ in linear terms, where $R_h$ is the event horizon, $R_i$ is the inner horizon, and $R_{\pm}$  represent cosmological horizons, one of which is negative, $R_-\in\mathbb{R}_{<0}$, and the other one is positive and bigger than the event horizon, $R_+>R_h$. In the following discussion, we fix $\Lambda=3$ and we work in the \emph{small black hole regime}, which corresponds to taking the black hole radius small compared to the norm of the de Sitter radius, $R_h\ll 1$.

Using the decomposition \eqref{decomposition}, the conformally coupled Klein-Gordon equation can be separated into an angular equation and a radial equation which read
\begin{equation}\label{ang-rad}
\begin{aligned}
&\frac{\mathrm{d}}{\mathrm{d}r}\biggl(\Delta_r(r)\frac{\mathrm{d}R(r)}{\mathrm{d}r}\biggr)+\Biggl[\frac{[\omega(r^2+a_{\mathrm{BH}}^2)-a_{\mathrm{BH}}m]^2\left(1+a_{\mathrm{BH}}^2\right)^2}{\Delta_r(r)}-2r^2-A_{\ell m}\Biggr]R(r)=0,\\
&\frac{\mathrm{d}}{\mathrm{d}x}\biggl[(a_{\mathrm{BH}}^2-x^2)\left(1+x^2\right)\frac{\mathrm{d}S(x)}{\mathrm{d}x}\biggr]+\Biggl[-\frac{\bigl\{\left(1+a_{\mathrm{BH}}^2\right)[\omega(a_{\mathrm{BH}}^2-x^2)-a_{\mathrm{BH}}m]\bigr\}^2}{(a_{\mathrm{BH}}^2-x^2)(1+x^2)}-2x^2+A_{\ell m }\Biggr]S(x)=0,
\end{aligned}
\end{equation}
where with $A_{\ell m}$ we denote the separation constant. Both equations can be written in Heun's form \cite{suzuki1998,PhysRevD.81.044005,Novaes:2018fry}. We first address the problem of quantization of the separation constant.

\subsection{Angular Problem}

The singularities of the angular equation are 
\begin{equation}
\pm a_{\mathrm{BH}},\pm i.
\end{equation}
The Kerr-de Sitter black hole solution is well defined if the $a_{\mathrm{BH}}$ parameter lies in the range $0<a_{\mathrm{BH}}<2-\sqrt{3}$. Indeed, the extreme cases in which two or more singularities coincide can be obtained by solving the system
\begin{equation}
\begin{cases}
\Delta_r(r)=0,\\
\Delta_r'(r)=0.
\end{cases}
\end{equation}
Solving the system in $r$ and $M$, gives the solutions
\begin{equation}
M=r-a_{\mathrm{BH}}^2 r-2 r^3,
\end{equation}
and 
\begin{equation}
r=\begin{cases}
\pm\sqrt{\frac{1-a_{\mathrm{BH}}^2-\sqrt{a_{\mathrm{BH}}^4-14 a_{\mathrm{BH}}^2+1}}{6}},\\
\pm\sqrt{\frac{1-a_{\mathrm{BH}}^2+\sqrt{a_{\mathrm{BH}}^4-14 a_{\mathrm{BH}}^2+1}}{6}}.
\end{cases}
\end{equation}
These are consistent with the physical requirements $M>0$ and $0<a_{\mathrm{BH}}<1$ if and only if $0<a_{\mathrm{BH}}<2-\sqrt{3}$.

Let us perform the following change of variables:
\begin{equation}
z=\frac{2i(x+a_{\mathrm{BH}})}{(a_{\mathrm{BH}}+i)(x+i)}.
\end{equation}
This change of variables maps
\begin{equation}
\begin{aligned}
\left(x_4=-i,x_1=-a_{\mathrm{BH}},x_2=i,x_3=a_{\mathrm{BH}},\infty\right)\mapsto \left(\infty,z_1=0,z_2=1,z_3:=\frac{4\,i\,a_{\mathrm{BH}}}{(a_{\mathrm{BH}}+i)^2},z_{\infty}:=\frac{2i}{a_{\mathrm{BH}}+i}\right).
\end{aligned}
\end{equation}
We note that, for $0<a_{\mathrm{BH}}<2-\sqrt{3}$, one has $|t|<1$.
Let us define 
\begin{equation}\label{angularquantities}
\begin{aligned}
\Delta_x(x)=&\,(a_{\mathrm{BH}}^2-x^2)\left(1+x^2\right),\\
\theta_k^{(a)}=&\,-\frac{(1+a_{\mathrm{BH}}^2)[\omega(a_{\mathrm{BH}}^2-x_k^2)-a_{\mathrm{BH}}m]}{\Delta'_x(x_k)},\ \ k=1,2,3.
\end{aligned}
\end{equation}
If we transform the angular wave function as
\begin{equation}
S(x)=(z-z_{\infty})\prod_{k=1}^3(z-z_i)^{-\theta_k^{(a)}}w(z),
\end{equation}
we can remove the apparent singularity in $z_{\infty}$ and the angular equation becomes a Heun equation
\begin{equation}\label{Heuneq}
\left[ \frac{\mathrm{d}^2 }{\mathrm{d}z^2}+\left( \frac{\gamma}{z}+\frac{\delta}{z-1}+\frac{\epsilon}{z-t} \right)\frac{\mathrm{d}}{\mathrm{d}z}+\frac{\alpha \beta z - q}{z(z-1)(z-t)} \right] w(z) = 0,
\end{equation}
with
\begin{equation}
\begin{aligned}
t&=\frac{4ia_{\mathrm{BH}}}{(a_{\mathrm{BH}}+i)^2},\\
\alpha&=1-i\omega+ia_{\mathrm{BH}}(m-a_{\mathrm{BH}}\omega),\quad\quad
\beta=1,\\
\gamma&=1-m,\quad\quad
\delta=1-i\omega+ia_{\mathrm{BH}}(m-a_{\mathrm{BH}}\omega),\quad\quad
\epsilon=1+m,\\
q&=\frac{A_{\ell m}+2a_{\mathrm{BH}} \left[\left(1-a_{\mathrm{BH}}^2\right) \omega + (a_{\mathrm{BH}}-i) m +i \right]}{(a_{\mathrm{BH}}+i)^2}.
\end{aligned}
\end{equation}
The dictionary that gives the parameter of the Heun's operator in normal form \eqref{heunnormalform}, is given by
\begin{equation}
\begin{aligned}
a_0=&\,\frac{m}{2},\quad a_t=-\frac{m}{2},\quad
a_1=\frac{i}{2}\left[\omega(1+a_{\mathrm{BH}}^2)-a_{\mathrm{BH}}m\right],\quad
a_{\infty}=-\frac{i}{2}\left[\omega(1+a_{\mathrm{BH}}^2)-a_{\mathrm{BH}}m\right],\\
u
=&\,\frac{1}{2(a_{\mathrm{BH}}-i)^2}\left[1+2A_{\ell m }-m^2+2a_{\mathrm{BH}}(i-im^2+2m\omega)+a_{\mathrm{BH}}^2(-1+8m-3m^2)+4a_{\mathrm{BH}}^3(m-2)\omega\right].
\end{aligned}
\end{equation}

We impose as boundary conditions the regularity of the solutions at $\theta=0,\pi$, which correspond to $x=\pm a_{\mathrm{BH}}$, and so to $z=t$ and $z= 0$.

For $x\sim a_{\mathrm{BH}}$ the original angular function has the following two behaviors:
\begin{equation}
\begin{aligned}
S_-^{(t)}(x)\sim(x-a_{\mathrm{BH}})^{\frac{m}{2}}\quad\quad
S_+^{(t)}(x)\sim(x-a_{\mathrm{BH}})^{-\frac{m}{2}}.
\end{aligned}
\end{equation}
Therefore, we take $S\sim S_-^{(t)}$ if $m\ge 0$, and we take $S\sim S_+^{(t)}$ if $m<0$.

For $x\sim -a_{\mathrm{BH}}$ the original angular function has the following two behaviors:
\begin{equation}
\begin{aligned}
S_-^{(0)}(x)\sim(x+a_{\mathrm{BH}})^{-\frac{m}{2}}\quad\quad
S_+^{(0)}(x)\sim(x+a_{\mathrm{BH}})^{\frac{m}{2}},
\end{aligned}
\end{equation}
Therefore, we take $S\sim S_-^{(0)}$ if $m\le 0$, and we take $S\sim S_+^{(0)}$ if $m>0$.

The boundary conditions are satisfied if the following requirement is imposed on the v.e.v. parameter $a$ (see Appendix~\ref{appendixNS} for the relevant definitions and conventions) which parameterizes the composite monodromy around $x=0$ and $x=t$:
\begin{equation}
a=\ell+\frac{1}{2},\ \ \text{with}\ \ \ell\ge |m|\ \ \text{and}\ \ell\in\mathbb{N}.
\end{equation}
The v.e.v. parameter $a$ is related to the parameter $u$ of the Heun differential equation through the \emph{Matone relation}
\begin{equation}
u=-\frac{1}{4}+a_t^2+a_0^2-a^2+t\frac{\partial F(t)}{\partial t},
\end{equation}
where $F(t)$ is the instanton partition function with four fundamental multiplets in the NS limit (see Appendix~\ref{appendixNS} for the relevant definitions and conventions).
Using the gravitational dictionary for $u$ and the quantization condition $a=\ell+\frac{1}{2}$, we obtain the following expansion of the separation constant:
\begin{equation}
\begin{aligned}
A_{\ell m s}=&\,(a_{\mathrm{BH}}-1)^2\ell(\ell+1)+2a_{\mathrm{BH}}m(a_{\mathrm{BH}}^2\omega-a_{\mathrm{BH}}m-\omega)-(a_{\mathrm{BH}}-1)^2t\frac{\partial F(t)}{\partial t}.
\end{aligned}
\end{equation}
As expected, expanding this expression around $a_{\mathrm{BH}}=0$ gives
\begin{equation}
\begin{aligned}
A_{\ell m s}=&\,\ell(\ell+1)-2\,m\, a_{\mathrm{BH}}\,\omega+\mathcal{O}(a_{\mathrm{BH}}^2).
\end{aligned}
\end{equation}

\subsection{Radial Problem and expression for the determinant}

For the radial equation, let us perform the following change of variables:
\begin{equation}
z=\frac{R_+-R_-}{R_+-R_h}\cdot\frac{r-R_h}{r-R_-}.
\end{equation}
This sends
\begin{equation}
\begin{aligned}
&(r_4=R_-,r_1=R_h,r_2=R_+,r_3=R_i,\infty)\mapsto \\
&\left(\infty,z_1=0,z_2=1,z_3:=t=\frac{R_+-R_-}{R_+-R_h}\cdot\frac{R_i-R_h}{R_i-R_-},z_{\infty}:=\frac{R_+-R_-}{R_+-R_h}\right).
\end{aligned}
\end{equation}
We remark that $t<0$, so that in the interval $z\in]0,1[$ there are no singularities.
Let us define 
\begin{equation}
\theta_k^{(r)}=\frac{i}{\Delta'_r(r_k)}[\omega(r_k^2+a_{\mathrm{BH}}^2)-a_{\mathrm{BH}}m](1+a_{\mathrm{BH}}^2),\quad k=1,\dots,4.
\end{equation}
If we transform the radial function as
\begin{equation}
R(r)=(z-z_{\infty})\prod_{k=1}^3(z-z_i)^{-\theta_k^{(r)}}w(z),
\end{equation}
we can remove the singularity in $z_{\infty}$ and the radial equation becomes a Heun equation \eqref{Heuneq}
with
\begin{equation}\label{KerrdSdictionary}
\begin{aligned}
t&=\frac{R_+-R_-}{R_+-R_h}\cdot\frac{R_i-R_h}{R_i-R_-},\\
\alpha&=1+\frac{2i}{\Delta'_r(R_-)}[\omega(R_-^2+a_{\mathrm{BH}}^2)-a_{\mathrm{BH}}m](1+a_{\mathrm{BH}}^2),\\
\beta&=1,\\
\gamma&=1-\frac{2i}{\Delta'_r(R_h)}[\omega(R_h^2+a_{\mathrm{BH}}^2)-a_{\mathrm{BH}}m](1+a_{\mathrm{BH}}^2),\\
\delta&=1-\frac{2i}{\Delta'_r(R_+)}[\omega(R_+^2+a_{\mathrm{BH}}^2)-a_{\mathrm{BH}}m](1+a_{\mathrm{BH}}^2),\\
\epsilon&=1-\frac{2i}{\Delta'_r(R_i)}[\omega(R_i^2+a_{\mathrm{BH}}^2)-a_{\mathrm{BH}}m](1+a_{\mathrm{BH}}^2),\\
q&= (t-1)(a_0+a_t)+t(a_1+a_t)-\frac{t(t-1)}{t-z_{\infty}}+t\,\alpha+\frac{2R_i^2+A_{\ell m}}{(R_h-R_+)(R_i-R_-)},
\end{aligned}
\end{equation}
where the indices of the singular points are
\begin{equation}\label{KerrdSdictionary2}
\begin{aligned}
a_0&=\frac{i}{\Delta'_r(R_h)}[\omega(R_h^2+a_{\mathrm{BH}}^2)-a_{\mathrm{BH}}m](1+a_{\mathrm{BH}}^2),\\
a_t&=\frac{i}{\Delta'_r(R_i)}[\omega(R_i^2+a_{\mathrm{BH}}^2)-a_{\mathrm{BH}}m](1+a_{\mathrm{BH}}^2),\\
a_1&=\frac{i}{\Delta'_r(R_+)}[\omega(R_+^2+a_{\mathrm{BH}}^2)-a_{\mathrm{BH}}m](1+a_{\mathrm{BH}}^2),\\
a_{\infty}&=\frac{i}{\Delta'_r(R_-)}[\omega(R_-^2+a_{\mathrm{BH}}^2)-a_{\mathrm{BH}}m](1+a_{\mathrm{BH}}^2),
\end{aligned}
\end{equation}
and the parameter $u$ in the Heun equation \eqref{heunnormalform} is given by
\begin{equation}\label{KerrdSdictionary3}
u=\frac{-2\,q+2\,t\,\alpha\,\beta+\gamma\,\epsilon-t\,(\gamma+\delta)\,\epsilon}{2(t-1)}.
\end{equation}

We distinguish two cases according to the sign of $\mathrm{Im}(\omega)$.

Let us start from the case $\mathrm{Im}(\omega)>0$. In this case $\mathrm{Re}(a_0)<0$ and $\mathrm{Re}(a_1)>0$. Then, the local solutions of the normal form of the Heun equation (and normalized as in \eqref{localsolutionsat0}) selected by the boundary conditions are 
\begin{equation}
\begin{aligned}
\psi_-^{(0)}(z)=&\,t^{-\epsilon/2}z^{\gamma/2}(z-1)^{\delta/2}(z-t)^{\epsilon/2}\mathrm{Heun}\left(t,q,\alpha,\beta,\gamma,\delta,z\right),\\
\psi_+^{(1)}(z)=&\,(1-t)^{-\epsilon/2}z^{\gamma/2}(z-1)^{1-\delta/2}(z-t)^{\epsilon/2}\times\\
&\times\left( \frac{z-t}{1-t} \right)^{- \alpha - 1 + \delta}\mathrm{Heun} \left( t, q - \alpha (\beta + \delta - 2) + (\delta - 1) (\alpha + \beta - 1 - t \gamma), \alpha + 1 - \delta, 1 + \gamma - \beta,2 - \delta, \gamma, t \frac{1-z}{t-z} \right).
\end{aligned}
\end{equation}
The connection formula between the two local solutions changes according to the position of the singularity $t$ in the $z$-space. 
The small black hole regime corresponds to the regime $|t|<1$\footnote{The other regime $|t|>1$, would lead to a simpler connection formula, more similar to a Hypergeometric-like connection problem, but still involving the presence of the NS functions (see Appendix~\ref{Heunconnectiont>1}).}.
The connection coefficient in terms of which we can express the determinant is the one in front of $\psi_{-}^{(1)}(z)$ starting from the solution $\psi_{-}^{(0)}(z)$ in the connection formula \eqref{connection0to1}:
\begin{equation}
\sum_{\theta'=\pm}\mathcal{M}_{-\theta'}(a_0,a;a_t)\mathcal{M}_{(-\theta')-}(a,a_1;a_{\infty})t^{-a_0+\theta' a}\exp(-\frac{1}{2}\partial_{a_0}F(t)+\frac{1}{2}\partial_{a_1}F(t)-\frac{\theta'}{2}\partial_aF(t)).
\end{equation}

In the case $\mathrm{Im}(\omega)<0$, the local (normalized) solutions selected by the boundary conditions are 
\begin{equation}
\begin{aligned}
&\psi_+^{(0)}(z)=e^{i\pi\left(-\delta/2-\epsilon/2\right)}t^{-\epsilon/2}z^{1-\gamma/2}(z-1)^{\delta/2}(z-t)^{\epsilon/2}\mathrm{Heun}\left(t,q-(\gamma-1)(t\,\delta+\epsilon),\alpha+1-\gamma,\beta+1-\gamma,2-\gamma,\delta,z\right),\\
&\psi_-^{(1)}(z)=(1-t)^{-\epsilon/2}z^{\gamma/2}(z-1)^{\delta/2}(z-t)^{\epsilon/2}\left( \frac{z-t}{1-t} \right)^{- \alpha} \mathrm{Heun} \left( t, q + \alpha (\delta - \beta), \alpha, \delta + \gamma - \beta, \delta, \gamma, t \frac{1-z}{t-z} \right).
\end{aligned}
\end{equation}
The connection coefficient in terms of which we can express the determinant is the one in front of $\psi_{+}^{(1)}(z)$ starting from the solution $\psi_{+}^{(0)}(z)$ in the connection formula \eqref{connection0to1}:
\begin{equation}
\sum_{\theta'=\pm}\mathcal{M}_{+\theta'}(a_0,a;a_t)\mathcal{M}_{(-\theta')+}(a,a_1;a_{\infty})t^{a_0+\theta' a}\exp(\frac{1}{2}\partial_{a_0}F(t)-\frac{1}{2}\partial_{a_1}F(t)-\frac{\theta'}{2}\partial_aF(t)).
\end{equation}

\subsection{Determinant of radial operator}

We can finally write the result for the determinant of the radial differential operator, following the procedure explained in Appendix~\ref{ratiodeterminants} and \ref{determinantofhyperg}. 
The reference problem we consider for the radial operator is a Hypergeometric problem having the same indices at the singular points $z=0$ and $z=1$.

For $\mathrm{Im}(\omega)>0$, we have $\mathrm{Re}(a_0)<0$ and $\mathrm{Re}(a_1)>0$. The formula for the (regularized) determinant reads
\begin{equation}\label{Imomega>0KerrdS}
\begin{aligned}
&\sum_{\theta'=\pm}\frac{2\pi\Gamma(-2\theta'a)\Gamma(1-2\theta'a)}{\prod_{\sigma=\pm}\Gamma\left(\frac{1}{2}-a_0-\theta'a+\sigma\,a_t\right)\Gamma\left(\frac{1}{2}-\theta'a+a_1+\sigma\,a_{\infty}\right)}t^{-a_0+\theta' a}\exp(-\frac{1}{2}\partial_{a_0}F(t)+\frac{1}{2}\partial_{a_1}F(t)-\frac{\theta'}{2}\partial_aF(t)).
\end{aligned}
\end{equation}

For $\mathrm{Im}(\omega)<0$, we have $\mathrm{Re}(a_0)>0$ and $\mathrm{Re}(a_1)<0$. The formula for the (regularized) determinant reads
\begin{equation}\label{Imomega<0KerrdS}
\begin{aligned}
&\sum_{\theta'=\pm}\frac{2\pi\Gamma(-2\theta'a)\Gamma(1-2\theta'a)}{\prod_{\sigma=\pm}\Gamma\left(\frac{1}{2}+a_0-\theta'a+\sigma\,a_t\right)\Gamma\left(\frac{1}{2}-\theta'a-a_1+\sigma\,a_{\infty}\right)}t^{a_0+\theta' a}\exp(\frac{1}{2}\partial_{a_0}F(t)-\frac{1}{2}\partial_{a_1}F(t)-\frac{\theta'}{2}\partial_aF(t)).
\end{aligned}
\end{equation}
We can summarize the two formulae together introducing $\eta=\mathrm{Im}(\omega)/|\mathrm{Im}(\omega)|$ as
\begin{equation}\label{KerrdSdeteta}
\begin{aligned}
\mathrm{det}(\mathcal{D}_{\mathrm{rad}}-A_{\ell m})=&\sum_{\theta'=\pm}\frac{2\pi\Gamma(-2\theta'a)\Gamma(1-2\theta'a)}{\prod_{\sigma=\pm}\Gamma\left(\frac{1}{2}-\eta a_0-\theta'a+\sigma\,a_t\right)\Gamma\left(\frac{1}{2}-\theta'a+\eta a_1+\sigma\,a_{\infty}\right)}\times\\
&\times t^{-\eta a_0+\theta' a}\exp(-\frac{\eta}{2}\partial_{a_0}F(t)+\frac{\eta}{2}\partial_{a_1}F(t)-\frac{\theta'}{2}\partial_aF(t)).
\end{aligned}
\end{equation}
The (anti-)quasinormal modes are directly given by the zeroes of the above expression.

\subsection{Schwarzschild-de Sitter spacetime in four dimensions}

In this subsection, we want to briefly comment on how the previous formula also gives the solution for the determinant of the same operator around the four-dimensional Schwarzschild-de Sitter black hole, which is a spherically symmetric spacetime. In particular, the angular problem, in this case, is solved by the spherical harmonics, and the only nontrivial problem is the radial one, which can be solved precisely as in the previous discussion, but with a simplified dictionary.

The metric describing the Schwarzschild-de Sitter black hole in four dimensions ($\rm SdS_4$) is
\begin{equation}\label{metric}
\mathrm{d} s^2=- f(r) \mathrm{d} t^2+f(r)^{-1} \mathrm{d} r^2+r^2 \mathrm{d} \Omega_{2}^2
\end{equation}
with 
\begin{equation} f(r)=1-\frac{2 M}{r}-\frac{\Lambda}{3}r^2,\end{equation}
where $M$ is the mass of the black hole and $\Lambda>0$ is the cosmological constant. In what follows, we fix $\Lambda=3$, therefore requiring $M$ to be in the range $0<M^2<1/27$, in order to have a physical solution with three real roots for the equation $rf(r)=0$. 
We denote these roots by 
\begin{equation} R_h, \quad R_{\pm},  \end{equation}
where $R_h\in \bigr]0,\frac{1}{\sqrt{3}}\bigl[$, the smallest positive real root, represents the event horizon, and $R_\pm$ are real and given in terms of $R_h$ by
\begin{equation}
R_{\pm}=\frac{-R_h\pm\sqrt{4-3R_h^2}}{2}.
\end{equation} 

The conformally coupled Klein-Gordon equation in the SdS geometry can be obtained from the Kerr-dS one by sending the rotation parameter $a_{\mathrm{BH}}\to 0$. This also sends the singularity $R_i\to 0$ and the angular equation becomes trivial, giving an exact result for the separation constant
 \begin{equation}
A_{\ell m}=\ell(\ell+1).
\end{equation}
The radial equation, instead, remains a Heun equation \eqref{heunnormalform}, whose parameters can be deduced from the ones in \eqref{KerrdSdictionary2} and \eqref{KerrdSdictionary3}:
\begin{equation}\label{SdSdictionary}
\begin{aligned}
t&=\frac{R_h}{R_-}\cdot\frac{R_+-R_-}{R_+-R_h},\quad\quad 
u=-\frac{2\ell(\ell+1)+(R_h+R_+)^2}{2R_+(2R_h+R_+)},\\
a_0&=\frac{i\omega R_h}{(R_h-R_-)(R_+-R_h)},\quad\quad
a_1=\frac{i\omega R_+}{(R_+-R_-)(R_h-R_+)},\quad\quad
a_t=0,\quad\quad
a_{\infty}=\frac{i\omega R_-}{(R_h-R_-)(R_--R_+)}.
\end{aligned}
\end{equation}
With this new dictionary, the expression of the determinant is given by \eqref{KerrdSdeteta}, as in the Kerr case.

\subsection{Pure de Sitter spacetime in four dimensions}

An additional simplification can be obtained from the previous problem in the limit in which $R_h\to 0$. This leads to the determinant of the same operator in the pure de Sitter spacetime in four dimensions. In this case, the radial problem reduces to a Hypergeometric differential equation:
\begin{equation}\label{puredeSittereq}
\left[\frac{\mathrm{d}^2}{\mathrm{d}z^2}+\frac{4\ell (\ell+1) (z-1)+\left(\omega ^2+1\right) z^2}{4 (z-1)^2 z^2}\right]\psi(z)=0.
\end{equation}
The indices of the singularities $z=0$ and $z=1$ are
\begin{equation}
a_0=-\ell-\frac{1}{2},\quad a_1=\frac{i\,\omega}{2}.
\end{equation}
The sign of $\mathrm{Re}(a_0)$ is always negative, whereas the sign of $\mathrm{Re}(a_1)$ depends on the sign of the imaginary part of the frequency. Therefore, the local solution selected around $z\sim 0$ is the one behaving like
\begin{equation}
\psi^{(0)}_-(z)=z^{\frac{1}{2}+\left(\ell+\frac{1}{2}\right)}\left(1+\mathcal{O}(z)\right).
\end{equation}
The selected solution around $z\sim 1$ is
\begin{equation}
\begin{aligned}
\psi^{(1)}_-(z)&=(z-1)^{\frac{1}{2}-\frac{i\omega}{2}}\left(1+\mathcal{O}(z-1)\right),\quad\text{if}\ \mathrm{Im}(\omega)>0,\\
\psi^{(1)}_+(z)&=(z-1)^{\frac{1}{2}+\frac{i\omega}{2}}\left(1+\mathcal{O}(z-1)\right),\quad\text{if}\ \mathrm{Im}(\omega)<0.
\end{aligned}
\end{equation}
Redefining the wave function as
\begin{equation}
\psi(z)=z^{\ell+1} (z-1)^{\frac{1}{2}-\frac{i\,\omega}{2}}w(z)
\end{equation}
we can rewrite the differential equation as in \eqref{Hypergeometricdiffeq} with
\begin{equation}
a=\ell+1,\quad b=\ell+1-i\omega,\quad c=2\ell+2.
\end{equation}
Using the connection formulae \eqref{hypergconnformulae} and the results in Appendix~\ref{determinantofhyperg}, the determinant can be written as
\begin{equation}\label{puredSdet}
\frac{2\pi}{\Gamma(\ell+1)\Gamma(\ell+1-i\,\eta\,\omega)},
\end{equation}
where $\eta=\mathrm{Im}(\omega)/|\mathrm{Im}(\omega)|$.
The zeros in $\omega$ of the previous functions are given by the quasinormal mode frequencies $\omega=-i(\ell+n+1)$ and by the anti-quasinormal mode frequencies $\omega=i(\ell+n+1)$.

\subsection{Reduction of the determinant from Schwarzschild-de Sitter to pure de Sitter}

In this subsection, we want to comment on how the result of the determinant in the pure de Sitter geometry can be obtained from the Schwarzschild-de Sitter case in the limit $R_h\to 0$ (or, equivalently, sending to zero the mass of the black hole $M\to 0$). We already stressed that starting from the determinant in the Kerr-de Sitter case and sending the rotation parameter $a_{\mathrm{BH}}\to 0$, one obtains the determinant of the Schwarzschild-de Sitter case, which has the same expression but with the reduced dictionary. This is a smooth limit, in the sense that the result can be obtained simply by looking at the limit of the parameters for $a_{\mathrm{BH}}\to 0$.
In the reduction to the pure de Sitter case, the equation becomes a Hypergeometric equation in a non-trivial way, namely by a collision of singularities.

Let us start by rewriting the determinant of the Schwarzschild-de Sitter case written in the following form:
\begin{equation}\label{connectionfromSdStodS}
\frac{2\pi\,\sum_{\theta'=\pm}\mathcal{M}_{-\theta'}(a_0,a;a_t)\mathcal{M}_{(-\theta')-}(a,a_1;a_{\infty})t^{-a_0+\theta' a}\exp(-\frac{1}{2}\partial_{a_0}F(t)+\frac{1}{2}\partial_{a_1}F(t)-\frac{\theta'}{2}\partial_aF(t))}{\Gamma(1-2a_0)\Gamma(2a_1)},
\end{equation}
where we took the $\mathrm{Im}(\omega)<0$ case (the $\mathrm{Im}(\omega)>0$ case is analogous).

By considering the limit $R_h\to 0$, we have to implement the limit $t\to 0$ in the Heun differential operator \eqref{heundiffop}. Comparing the reduced operator with \eqref{puredeSittereq}, we can see that the new index $\tilde{a}_0$ at $z=0$ is given by
\begin{equation}
\tilde{a}_0^2=-\frac{1}{4}-u+a_0^2+a_t^2\bigg|_{R_h\to 0}=\frac{(2\ell+1)^2}{4}.
\end{equation}
Note that this is not obtained smoothly from $a_0$ by sending $R_h\to 0$ because of the collision of singularities.
Moreover,
\begin{equation}
a^2\bigg|_{R_h\to 0}=-\frac{1}{4}-u+a_0^2+a_t^2=\tilde{a}_0^2.
\end{equation}
Indeed, when $0$ and $t$ collide, the monodromy parametrized by $a$ becomes simply the monodromy around $z=0$. Therefore, in \eqref{connectionfromSdStodS}, the first connection matrix $\mathcal{M}_{-\theta'}(a_0,a;a_t)$ trivializes and reduces to the identity matrix\footnote{This can also be seen from the Liouville three-point functions by considering one of the three insertions to reduce to the identity insertion, see Appendix~A in \cite{Bonelli:2022ten} for the detailed definitions and conventions.}.

Let us now fix, consistently with the previous subsections, the signs $\tilde{a}_0=-\ell-\frac{1}{2}$ and $a\to \ell+\frac{1}{2}$. Then, the determinant \eqref{connectionfromSdStodS}, in the limit $R_h\to 0$, reduces to
\begin{equation}
\begin{aligned}
&\frac{2\pi}{\Gamma(2\ell+2)\Gamma(i\,\omega)}\left[\sum_{\theta'=\pm}\frac{\Gamma\left(1-2\theta'\,\left(\ell+\frac{1}{2}\right)+\mathcal{O}\left(R_h\right)\right)\Gamma\left(i\,\omega\right)}{\Gamma\left(\frac{1}{2}-\theta'\,\left(\ell+\frac{1}{2}\right)+\mathcal{O}\left(R_h\right)\right)\Gamma\left(\frac{1}{2}-\theta'\,\left(\ell+\frac{1}{2}\right)+i\,\omega\right)}(-2R_h)^{\ell+\frac{1}{2}+\theta'\left(\ell+\frac{1}{2}\right)}\right]\cdot\left[1+\mathcal{O}\left(R_h\log(R_h)\right)\right]=\\
&=\frac{2\pi}{\Gamma(2\ell+2)}\left[\frac{\Gamma\left(2\ell+2\right)}{\Gamma\left(\ell+1\right)\Gamma\left(\ell+1+i\,\omega\right)}+\frac{\Gamma\left(-2\ell+\mathcal{O}\left(R_h\right)\right)}{\Gamma\left(-\ell+\mathcal{O}\left(R_h\right)\right)\Gamma\left(-\ell+i\,\omega\right)}(-2R_h)^{2\ell+1}\right]\cdot\left[1+\mathcal{O}\left(R_h\log(R_h)\right)\right]=\\
&=\frac{2\pi}{\Gamma\left(\ell+1\right)\Gamma\left(\ell+1+i\,\omega\right)}\left[1+\mathcal{O}\left(R_h\log(R_h)\right)\right],
\end{aligned}
\end{equation}
which is the result obtained in the previous subsection, \eqref{puredSdet}.
Passing to the final result, we used the fact that the choice of the sign $\theta'=+$ forces the corresponding channel to go to zero, as can be seen from the dependence on $R_h^{2\ell+1}$ and noticing that the ratio of Gamma functions $\Gamma\left(-2\ell+\mathcal{O}\left(R_h\right)\right)/\Gamma\left(-\ell+\mathcal{O}\left(R_h\right)\right)$ gives a finite quantity in the $R_h\to 0$ limit.

\section{Schwarzschild anti-de Sitter spacetime in five dimensions}\label{section4}

The metric of the five-dimensional Schwarzschild-anti-de Sitter black hole (SAdS$_5$) is
\begin{equation}
\mathrm{d}s^2=-f(r)\mathrm{d}t^2+f(r)^{-1}\mathrm{d}r^2+r^2\mathrm{d}\Omega_3^2,
\end{equation}
where $\mathrm{d}\Omega_3^2$ is the volume element of the 3-sphere and, normalizing the AdS radius to 1, 
\begin{equation}
f(r)=\left(1-\frac{R_h^2}{r^2}\right)\left(r^2+R_h^2+1\right),
\end{equation}
where $R_h$ is the radius of the black hole horizon. We again work in the small black hole regime, $0<R_h\ll 1$.

The wave equation satisfied by (the Fourier modes $\phi_{\ell,\omega}$ of) a massive scalar field $\Phi$ in this black hole background is given by 
\begin{equation}
\left[\frac{1}{r^3}\frac{\mathrm{d}}{\mathrm{d}r}\left(r^3f(r)\frac{\mathrm{d}}{\mathrm{d}r}\right)+\frac{\omega^2}{f(r)}-\frac{\ell(\ell+2)}{r^2}-\Delta(\Delta-4)\right]\phi_{\ell,\omega}(r)=0,
\end{equation}
where $\Delta$ is the dimension of the scalar-operator dual to the scalar field in the bulk, related to the mass $\mu$ of the field by $\mu=\sqrt{\Delta(\Delta-4)}$. The problem is symmetric under $\Delta\mapsto 4-\Delta$. We assume in what follows $\Delta>2$ and $\Delta\notin\mathbb{N}$ in order not to be in a log case.

Defining a new variable
\begin{equation}
z=\frac{r^2-R_h^2}{r^2+R_h^2+1},
\end{equation}
and redefining the wave function as
\begin{equation}
\phi_{\ell,\omega}(r)=(z-1)^{2-\frac{\Delta }{2}} z^{-\frac{i \omega R_h}{4 R_h^2+2}}w_{\ell,\omega}(z),
\end{equation}
where
\begin{equation}
\begin{aligned}
t=\,-\frac{R_h^2}{R_h^2+1},\quad \gamma=\,1-\frac{i\omega R_h}{2R_h^2+1},\quad \delta=\,3-\Delta,\quad \epsilon=\,1,
\end{aligned}
\end{equation}
the differential equation becomes a Heun equation \eqref{Heuneq}, where the complete dictionary is given by
\begin{equation}
\begin{aligned}
t=&\,-\frac{R_h^2}{R_h^2+1},\\
\alpha=&\,\frac{(4-\Delta ) \left(2 R_h^2+1\right)+\omega\sqrt{R_h^2+1}  -i R_h \omega }{4 R_h^2+2},\\
\beta=&\,\frac{(4-\Delta )(1+2R_h^2)  -\omega\sqrt{R_h^2+1}  -i R_h \omega}{4 R_h^2+2},\\
\gamma=&\,1-\frac{i\omega R_h}{2R_h^2+1},\quad \delta=\,3-\Delta,\quad \epsilon=\,1,\\
q=&\,-\frac{\left(2 R_h^2+1\right) \left(\ell (\ell+2)+(\Delta -4) (\Delta -2) R_h^2\right)+2 i R_h \omega  \left((\Delta -2) R_h^2+1\right)-R_h^2 \omega ^2}{8 R_h^4+12 R_h^2+4}.
\end{aligned}
\end{equation}
We remark that again $t$ is real and negative.
The parameters of the equation in normal form \eqref{heunnormalform} are
\begin{equation}\label{SAdSdictionary}
\begin{aligned}
a_0=&\,\frac{i\omega R_h}{2(2R_h^2+1)},\quad a_t=\,0,\quad a_1=\,\frac{\Delta-2}{2},\quad a_{\infty}=\,\frac{\omega \sqrt{R_h^2+1}}{2(2R_h^2+1)},\quad u=\,-\frac{\ell(\ell+2)+2 R_h^2+2}{8 R_h^2+4}.
\end{aligned}
\end{equation}
In the $z$ variable, the black hole horizon is located at $z=0$ and the AdS boundary at $z=1$.
The main important difference in this case with respect to the problems in an asymptotically de Sitter spacetime is the fact that the choice of the local solution near the AdS boundary does not depend on the sign of the imaginary part of $\omega$, but just on the parameter $\Delta$. This is because the corresponding boundary condition is imposed at the AdS boundary and not in a horizon of the geometry. With the assumptions we made on $\Delta$, the local solution of the Heun equation in normal form selected at $z=1$ is given by
\begin{equation}
\begin{aligned}
\psi_+^{(1)}(z)=&\,(1-t)^{-\epsilon/2}z^{\gamma/2}(z-1)^{1-\frac{\delta}{2}}(z-t)^{\epsilon/2}\times\\
&\times\left(\frac{z-t}{1-t}\right)^{-\alpha-1+\delta}\mathrm{Heun}\left(t,q-(\delta-1)\gamma t-(\beta-1)(\alpha-\delta+1),-\beta+\gamma+1,\alpha-\delta+1,2-\delta,\gamma,t\frac{1-z}{t-z}\right).
\end{aligned}
\end{equation}
For the choice of the local solution around $r=R_h$, we again divide the cases according to the sign of $\mathrm{Im}(\omega)$.
In the case $\mathrm{Im}(\omega)>0$ we have $\mathrm{Re}(a_0)<0$ and the local solution of the normal form of the Heun equation (and normalized as in \eqref{localsolutionsat0}) selected by the boundary condition is 
\begin{equation}
\begin{aligned}
\psi_-^{(0)}(z)=&\,t^{-\epsilon/2}z^{\gamma/2}(z-1)^{\delta/2}(z-t)^{\epsilon/2}\mathrm{Heun}\left(t,q,\alpha,\beta,\gamma,\delta,z\right).
\end{aligned}
\end{equation}
Considering again the regime in which $|t|<1$, the connection coefficient in terms of which we can express the determinant is the one in front of $\psi_{-}^{(1)}(z)$ starting from the solution $\psi_{-}^{(0)}(z)$ in the connection formula \eqref{connection0to1}, namely
\begin{equation}
\sum_{\theta'=\pm}\mathcal{M}_{-\theta'}(a_0,a;a_t)\mathcal{M}_{(-\theta')-}(a,a_1;a_{\infty})t^{-a_0+\theta' a}\exp(-\frac{1}{2}\partial_{a_0}F(t)+\frac{1}{2}\partial_{a_1}F(t)-\frac{\theta'}{2}\partial_aF(t)).
\end{equation}

In the case $\mathrm{Im}(\omega)<0$, the local (normalized) solution around $z=0$ selected by the boundary condition is 
\begin{equation}
\begin{aligned}
&\psi_+^{(0)}(z)=e^{i\pi\left(-\delta/2-\epsilon/2\right)}t^{-\epsilon/2}z^{1-\gamma/2}(z-1)^{\delta/2}(z-t)^{\epsilon/2}\mathrm{Heun}\left(t,q-(\gamma-1)(t\,\delta+\epsilon),\alpha+1-\gamma,\beta+1-\gamma,2-\gamma,\delta,z\right).
\end{aligned}
\end{equation}
The connection coefficient in terms of which we can express the determinant is the one in front of $\psi_{-}^{(1)}(z)$ starting from the solution $\psi_{+}^{(0)}(z)$ in the connection formula \eqref{connection0to1}, that is
\begin{equation}
\sum_{\theta'=\pm}\mathcal{M}_{+\theta'}(a_0,a;a_t)\mathcal{M}_{(-\theta')-}(a,a_1;a_{\infty})t^{a_0+\theta' a}\exp(\frac{1}{2}\partial_{a_0}F(t)+\frac{1}{2}\partial_{a_1}F(t)-\frac{\theta'}{2}\partial_aF(t)).
\end{equation}

\subsection{Determinant of radial operator}

We can again write the expression of the (regularized) determinant according to the sign of $\mathrm{Im}(\omega)$, remembering that we always have $\mathrm{Re}(a_1)>0$.

For $\mathrm{Im}(\omega)>0$, we have $\mathrm{Re}(a_0)<0$ and the formula for the (regularized) determinant reads
\begin{equation}
\begin{aligned}
&\sum_{\theta'=\pm}\frac{2\pi\Gamma(-2\theta'a)\Gamma(1-2\theta'a)}{\prod_{\sigma=\pm}\Gamma\left(\frac{1}{2}-a_0-\theta'a+\sigma\,a_t\right)\Gamma\left(\frac{1}{2}-\theta'a+a_1+\sigma\,a_{\infty}\right)}t^{-a_0+\theta' a}\exp(-\frac{1}{2}\partial_{a_0}F(t)+\frac{1}{2}\partial_{a_1}F(t)-\frac{\theta'}{2}\partial_aF(t)).
\end{aligned}
\end{equation}

For $\mathrm{Im}(\omega)<0$, we have $\mathrm{Re}(a_0)>0$ and the formula for the (regularized) determinant reads
\begin{equation}
\begin{aligned}
&\sum_{\theta'=\pm}\frac{2\pi\Gamma(-2\theta'a)\Gamma(1-2\theta'a)}{\prod_{\sigma=\pm}\Gamma\left(\frac{1}{2}+a_0-\theta'a+\sigma\,a_t\right)\Gamma\left(\frac{1}{2}-\theta'a+a_1+\sigma\,a_{\infty}\right)}t^{a_0+\theta' a}\exp(\frac{1}{2}\partial_{a_0}F(t)+\frac{1}{2}\partial_{a_1}F(t)-\frac{\theta'}{2}\partial_aF(t)).
\end{aligned}
\end{equation}

We can unify the two formulae above by introducing $\eta=\mathrm{Im}(\omega)/|\mathrm{Im}(\omega)|$
and get
\begin{equation}\label{5dSads}
\begin{aligned}
\mathrm{det}(\mathcal{D}_{\mathrm{rad}}-A_{\ell m})=&\sum_{\theta'=\pm}\frac{2\pi\Gamma(-2\theta'a)\Gamma(1-2\theta'a)}{\prod_{\sigma=\pm}\Gamma\left(\frac{1}{2}-\eta a_0-\theta'a+\sigma\,a_t\right)\Gamma\left(\frac{1}{2}-\theta'a+a_1+\sigma\,a_{\infty}\right)}\times\\
&\times t^{-\eta a_0+\theta' a}\exp(-\frac{\eta}{2}\partial_{a_0}F(t)+\frac{1}{2}\partial_{a_1}F(t)-\frac{\theta'}{2}\partial_aF(t)).
\end{aligned}
\end{equation}

\subsection{Pure anti-de Sitter spacetime in five dimensions}

As in the asymptotically de Sitter case, sending $R_h\to 0$ simplifies the problem, reducing it to the pure $\mathrm{AdS}_5$ case, whose relevant differential equation is again of Hypergeometric type.
Following the same procedure of the $\mathrm{SAdS}_5$ problem, or reducing the dictionary in the limit $R_h=0$, we can write the radial differential equation in normal form as
\begin{equation}
\psi''(z)+\frac{\ell(\ell+2)(z-1)+z \left[\omega ^2 (1-z)+z-(\Delta -2)^2\right]}{4 (z-1)^2 z^2}\,\psi(z)=0,
\end{equation}
which is a Hypergeometric differential equation.

The boundary conditions are imposed at the origin $z=0$ and at the AdS boundary $z=1$. We notice that both points do not represent horizons of the geometry, and the indices in these singular points do not depend on $\omega$: 
\begin{equation}
a_0=-\frac{\ell+1}{2},\quad a_1=1-\frac{\Delta }{2}.
\end{equation}
Assuming again $\Delta>2$ and $\Delta\notin\mathbb{N}$, the determinant can be written as
\begin{equation}
\frac{2\pi}{\Gamma\left(\frac{\Delta +\ell-\omega}{2}\right)\Gamma\left(\frac{\Delta +\ell+\omega}{2}\right)}.
\end{equation}
We notice that the zeros in $\omega$ of the determinant are real and given by the normal modes of AdS$_5$
\begin{equation}\label{AdSnormalmodes}
\omega=-\ell-\Delta-2n\quad \text{and}\quad \omega=\ell+\Delta+2n,\quad \text{with}\ \  n\in\mathbb{Z}_{\ge 0}. 
\end{equation}


\section{Detailed analysis of the effective actions}\label{conclusion}

In the previous sections, we computed the determinant of the radial differential operator in Heun's form, which can be written as
\begin{equation}
\mathrm{det}\left(\mathcal{D}_{\mathrm{rad}}-A_{\ell m}\right)=2\pi\frac{\mathcal{C}_{12}}{\Gamma(1+2\theta_0a_0)\Gamma(2\theta_1a_1)},
\end{equation}
where $a_0,a_1$ denote the indices of the singularities at $z=0$ and $z=1$ of the Heun differential operator, $\mathcal{C}_{12}$ denotes the Heun connection coefficient between the local solution at $z=0$ satisfying the boundary condition and the discarded local solution at $z=1$, and where $\theta_0,\theta_1=\pm$, according to the sign of $\mathrm{Im}(\omega)$.

However, it is important to notice that the problems we considered are PT-symmetric, and the full determinants \eqref{fulldet} are symmetric for the transformation $\omega\mapsto-\omega$. In particular, the contribution coming from the analytic continuation for $\mathrm{Im}(\omega)<0$ gives the same result obtained for the analytic continuation for $\mathrm{Im}(\omega)>0$. More precisely, our determinant for the radial part has the following property:
\begin{equation}
\mathrm{det}\left(\mathcal{D}_{\mathrm{rad}}-A_{\ell \vec{m}}\right)[\omega]=\begin{cases}
\mathrm{det}^{(+)}\left(\mathcal{D}_{\mathrm{rad}}-A_{\ell \vec{m}}\right)[\omega],\quad\text{for}\quad\mathrm{Im}(\omega)>0,\\
\mathrm{det}^{(-)}\left(\mathcal{D}_{\mathrm{rad}}-A_{\ell \vec{m}}\right)[\omega],\quad\text{for}\quad\mathrm{Im}(\omega)<0,
\end{cases}
\end{equation}
with
\begin{equation}
\mathrm{det}^{(-)}\left(\mathcal{D}_{\mathrm{rad}}-A_{\ell \vec{m}}\right)[\omega]=\mathrm{det}^{(+)}\left(\mathcal{D}_{\mathrm{rad}}-A_{\ell \vec{m}}\right)[-\omega].
\end{equation}

We conclude that our final result for the one-loop effective action of a real scalar field is given by
\begin{equation}
\log(\mathrm{det}\left(\Box-\mu^2\right))=\int_{-\infty}^{\infty}\mathrm{d}\omega\sum_{\ell,\vec{m}}{\log}
\left(
\mathrm{det}^{(+)}(\mathcal{D}_{\mathrm{rad}}-A_{\ell \vec{m}})[\omega]
\right).
\end{equation}
In the above formula, one has to substitute
\begin{equation}\label{finalresult}
\begin{aligned}
\mathrm{det}^{(+)}(\mathcal{D}_{\mathrm{rad}}-A_{\ell m})[\omega]=&\sum_{\theta'=\pm}\frac{2\pi\Gamma(-2\theta'a)\Gamma(1-2\theta'a)}{\prod_{\sigma=\pm}\Gamma\left(\frac{1}{2}- a_0-\theta'a+\sigma\,a_t\right)\Gamma\left(\frac{1}{2}-\theta'a+ a_1+\sigma\,a_{\infty}\right)}\times\\
&\times t^{- a_0+\theta' a}\exp(-\frac{1}{2}\partial_{a_0}F(t)+\frac{1}{2}\partial_{a_1}F(t)-\frac{\theta'}{2}\partial_aF(t)),
\end{aligned}
\end{equation}
where, for the analyzed problems, the dictionaries of the quantities are given in \eqref{KerrdSdictionary2} for the Kerr-de Sitter case, in \eqref{SdSdictionary} for the Schwarzschild-de Sitter case, and in \eqref{SAdSdictionary} for the Schwarzschild-anti-de Sitter case.

One can see that the two summands in \eqref{finalresult} have different behaviors, which are determined by the exponential factor $t^{\theta'\,a}$. 

Since we always took $a$ to be positive, in the limit in which $t$ is small (that in our problems corresponded to the small black hole regime) we can argue that the term proportional to $t^{a}$ is subleading compared to the one proportional to $t^{-a}$. In particular, we can write
\begin{equation}\label{conncoeffdecomposition}
\begin{aligned}
&\log\left(\mathrm{det}^{(+)}(\mathcal{D}_{\mathrm{rad}}-A_{\ell m})[\omega]\right)=\\
&=\,\log\left(\frac{2\pi\Gamma(2a)\Gamma(1+2a)}{\prod_{\sigma=\pm}\Gamma\left(\frac{1}{2}- a_0+a+\sigma\,a_t\right)\Gamma\left(\frac{1}{2}+a+ a_1+\sigma\,a_{\infty}\right)} t^{- a_0-a}\exp(-\frac{1}{2}\partial_{a_0}F(t)+\frac{1}{2}\partial_{a_1}F(t)+\frac{1}{2}\partial_aF(t))\right)+\\
&\ \ \ \ +\log\left(1-\frac{\Gamma(-2a)^2}{\Gamma(2a)^2}\prod_{\sigma=\pm}\frac{\Gamma\left(\frac{1}{2}-a_0+a+\sigma\,a_t\right)\Gamma\left(\frac{1}{2}+a+a_1+\sigma\,a_{\infty}\right)}{\Gamma\left(\frac{1}{2}-a_0-a+\sigma\,a_t\right)\Gamma\left(\frac{1}{2}-a+a_1+\sigma\,a_{\infty}\right)}t^{2a}\exp\bigl(-\partial_aF(t)\bigr)\right),
\end{aligned}
\end{equation}
where the second line encodes the correction terms to the leading result in the first line.

This suggests that, in the decomposition in \eqref{finalresult}, the effects due to the presence of the black hole are subleading compared to the ones due to the asymptotic geometry. This is equivalent to saying that the contribution to the near-horizon zone is subleading compared to the far-zone (for a discussion on the distinction of these regions see \cite{Bautista:2023sdf}).

We add that, since the leading order of $a$ in the small black hole regime is determined by the angular quantum number $\ell$, the previous decomposition is also significant in the limit $\ell\gg 1$. Indeed, for large values of $\ell$, the term $t^{2a}$ is exponentially suppressed, and the first line of the previous decomposition already gives a good estimate for the (logarithm of the) radial determinant.

We finally remark that, in the Schwarzschild-de Sitter case, this small $R_h$ expansion gives purely imaginary QNMs (see \cite{Aminov:2023jve} and \cite{Jansen:2017oag}), as it happens for pure de Sitter spacetime. Analogously, in the Schwarzschild-anti-de Sitter case, neglecting the second channel in \eqref{finalresult}, produces purely real QNMs (see \cite{Dodelson:2022yvn}), as it happens for pure anti-de Sitter spacetime. 

More precisely, in the small $R_h$ expansion of the QNMs, 
\begin{equation}\label{QNMexpansion}
\omega=\sum_{k\ge 0}\omega_{k}\,R_h^k,
\end{equation}
where the dependence on the quantum numbers is implied, the first orders $\omega_{k}$ can be found by looking at the zeros of
\begin{equation}
\frac{\Gamma(2a)\Gamma(1+2a)}{\prod_{\sigma=\pm}\Gamma\left(\frac{1}{2}- a_0+a+\sigma\,a_t\right)\Gamma\left(\frac{1}{2}+a+ a_1+\sigma\,a_{\infty}\right)} t^{- a_0-a}\exp(-\frac{1}{2}\partial_{a_0}F(t)+\frac{1}{2}\partial_{a_1}F(t)+\frac{1}{2}\partial_aF(t)),
\end{equation}
which is equivalent to looking at the poles in the Gamma functions in the denominator\footnote{This is justified by gauge theory considerations, since $F(t)$ can be expressed as a series expansion in $t$ (see Appendix~\ref{appendixNS}), and, therefore, there are no zeroes in the exponential functions.}. 
For the four-dimensional Schwarzschild-de Sitter case this gives the correct coefficients $\omega_{k}$ for $0\le k\le 2\ell+1$ (see \cite{Aminov:2023jve}), while for the five-dimensional Schwarzschild-anti-de Sitter case this gives the correct coefficients $\omega_{k}$ for $0\le k\le 2\ell+2$ (see \cite{Dodelson:2022yvn}\footnote{In \cite{Dodelson:2022yvn} the small expansion parameter is $\mu$ which in the small black hole regime behaves like $\mu\sim R_h^2$. The near-horizon zone starts contributing when the QNMs develop an imaginary part, which behaves like $\mu^{\ell+\frac{3}{2}}$.}).

For the higher-order coefficients $\omega_k$, the quantization condition involves both channels of the connection coefficient:
\begin{equation}
\begin{aligned}
&\frac{\Gamma(2a)\Gamma(1+2a)}{\prod_{\sigma=\pm}\Gamma\left(\frac{1}{2}- a_0+a+\sigma\,a_t\right)\Gamma\left(\frac{1}{2}+a+ a_1+\sigma\,a_{\infty}\right)} t^{- a_0-a}\exp(-\frac{1}{2}\partial_{a_0}F(t)+\frac{1}{2}\partial_{a_1}F(t)+\frac{1}{2}\partial_aF(t))+\\
&\frac{\Gamma(-2a)\Gamma(1-2a)}{\prod_{\sigma=\pm}\Gamma\left(\frac{1}{2}- a_0-a+\sigma\,a_t\right)\Gamma\left(\frac{1}{2}-a+ a_1+\sigma\,a_{\infty}\right)} t^{- a_0+a}\exp(-\frac{1}{2}\partial_{a_0}F(t)+\frac{1}{2}\partial_{a_1}F(t)-\frac{1}{2}\partial_aF(t))=0,
\end{aligned}
\end{equation}
that is
\begin{equation}\label{quantBperiod}
\frac{\Gamma(-2a)^2\prod_{\sigma=\pm}\Gamma\left(\frac{1}{2}- a_0+a+\sigma\,a_t\right)\Gamma\left(\frac{1}{2}+a+ a_1+\sigma\,a_{\infty}\right)}{\Gamma(2a)^2\prod_{\sigma=\pm}\Gamma\left(\frac{1}{2}- a_0-a+\sigma\,a_t\right)\Gamma\left(\frac{1}{2}-a+ a_1+\sigma\,a_{\infty}\right)} t^{2a}\exp(-\partial_aF(t))=1.
\end{equation}
Again, this is a manifestation of the fact that in the small $R_h$ regime the contribution of the near-horizon zone is delayed compared to the far-zone, and the order of delay is determined by the angular quantum number $\ell$.

\subsection{Wick rotation and the thermal version}

In this final part, we 
analyze the thermal version of the one-loop quantum effective actions,
show how our results generalize the ones already present in the literature \cite{Denef:2009kn,Law:2022zdq} and reduce to the latter when the relevant differential equation reduces to the Hypergeometric one. 

Let us Wick rotate the spacetime metric to real-time by defining $t = i\tau$, where $\tau$ has periodicity equal to the inverse of the temperature $T$ of the spacetime. We can introduce the thermal frequencies by setting 
\begin{equation}\label{thermalfrequancies}
i\,\omega_k=2\,\pi\, T\,k,\quad k\in\mathbb{Z}.
\end{equation}
With these redefinitions, it is possible to connect our results with the one in \cite{Denef:2009kn}. In particular, the results for $\omega$ with a positive imaginary part correspond to computation with $k<0$, whereas the results for $\omega$ with a negative imaginary part correspond to computation with $k>0$.

Let us see the match in the pure de Sitter and anti-de Sitter cases, where the radial differential equations are of Hypergeometric type. 

In the four-dimensional de Sitter case, our result, using also the PT symmetry, can be rewritten as\footnote{In writing the equality, we neglect UV divergencies due to the infinite products over the quantum numbers in the RHS. These should be cured by subtracting local counterterms, which can be analyzed, for example, with heat kernel methods or WKB-type approximations. In \cite{Law:2022zdq}, the authors also comment on the possibility of absorbing these divergences into the cosmological constant, Newton's constant, and local couplings to higher curvature terms in the gravity sector.\label{footnoteprod}}
\begin{equation}
\log\left(\mathrm{det}\left(\Box-\mu^2\right)\right)=\sum_{k\in\mathbb{Z}}\sum_{\ell,m}\log\left(\frac{2\pi}{\Gamma\left(\ell+1\right)\Gamma\left(\ell+1+2\pi\,T\,|k|\right)}\right)=\sum_{k\in\mathbb{Z}}\sum_{\ell=0}^{\infty}(2\ell+1)\log\left(\frac{2\pi}{\ell!\,\Gamma\left(\ell+1+2\pi\,T\,|k|\right)}\right),
\end{equation}
where we used that the degeneracy for each $\ell\ge 0$ is equal to $2\ell+1$ due to spherical symmetry.

Using Weierstrass's definition of the Gamma function
\begin{equation}
\begin{aligned}
\Gamma\left(\ell+1+2\pi\,T\,|k|\right)=\frac{\exp(-\gamma\left(\ell+1+2\pi\,T\,|k|\right))}{\ell+1+2\pi\,T\,|k|}\prod_{n=1}^{\infty}\left[\left(1+\frac{\ell+1+2\pi\,T\,|k|}{n}\right)^{-1}\exp(\frac{\ell+1+2\pi\,T\,|k|}{n})\right],
\end{aligned}
\end{equation}
%
%
we can write the full determinant as (for the equality in the formula see the comment in footnote \ref{footnoteprod})
\begin{equation}\label{fulldetpuredS4}
\begin{aligned}
\mathrm{det}\left(\Box-\mu^2\right)&=\prod_{k\in\mathbb{Z}}\prod_{\ell=0}^{\infty}\left[\frac{2\pi(\ell+1+2\,\pi\,T\,|k|)\prod_{n=1}^{\infty}\left(1+\frac{\ell+1+2\pi\,T\,|k|}{n}\right)}{\ell!\,\exp(-\gamma\left(\ell+1+2\pi\,T\,|k|\right))\prod_{n=1}^{\infty}\exp(\frac{\ell+1+2\pi\,T\,|k|}{n})}\right]^{2\ell+1}=\\
&=\prod_{k\in\mathbb{Z}}\prod_{\ell=0}^{\infty}\left[\frac{2\pi(2\,\pi\,T)\prod_{n=1}^{\infty}\frac{2\,\pi\,T}{n}\prod_{n=0}^{\infty}\left(|k|+\frac{\ell+n+1}{2\,\pi\,T}\right)}{\ell!\,\exp(-\gamma\left(\ell+1+2\pi\,T\,|k|\right))\prod_{n=1}^{\infty}\exp(\frac{\ell+1+2\pi\,T\,|k|}{n})}\right]^{2\ell+1}=\\
&=\prod_{k\in\mathbb{Z}}\prod_{\ell=0}^{\infty}\left[\frac{2\pi}{\ell!\,\exp(-\gamma\left(\ell+1+2\pi\,T\,|k|\right))}\frac{\prod_{n=0}^{\infty}(2\,\pi\,T)}{\prod_{n=1}^{\infty}n\,\exp(\frac{\ell+1+2\pi\,T\,|k|}{n})}\right]^{2\ell+1}\times\\
&\ \ \times\prod_{k\in\mathbb{Z}}\prod_{\ell=0}^{\infty}\prod_{n=0}^{\infty}\left(|k|+\frac{\ell+n+1}{2\,\pi\,T}\right)^{2\ell+1},
\end{aligned}
\end{equation}
where, in the final result, the first line is just an overall entire function (without any poles and zeroes in $k$, which corresponds to $\omega$). Notice that since the QNMs of pure de Sitter spacetime are given by $-i(\ell+n+1)$, this result is consistent with formula (2.10) in \cite{Law:2022zdq}.

In the pure AdS$_5$ case, the reasoning is analogous and, up to the overall factor, the structure of zeros can be seen from the infinite product arising from the Gamma functions
\begin{equation}
\frac{1}{\Gamma\left(\frac{\Delta+\ell-\omega_k}{2}\right)\Gamma\left(\frac{\Delta+\ell+\omega_k}{2}\right)}=\frac{1}{\Gamma\left(\frac{\Delta+\ell+2\,\pi\,i\,T\,|k|}{2}\right)\Gamma\left(\frac{\Delta+\ell-2\,\pi\,i\,T\,|k|}{2}\right)}.
\end{equation}
The infinite product contribution gives
\begin{equation}\label{fulldetpureAdS5}
\begin{aligned}
\mathrm{det}\left(\Box-\mu^2\right)\sim&\,\prod_{k\in\mathbb{Z}}\prod_{\ell=0}^{\infty}\left[\frac{\Delta+\ell+2\,\pi\,i\,T\,|k|}{2}\frac{\Delta+\ell-2\,\pi\,i\,T\,|k|}{2}\prod_{n=1}^{\infty}\left(1+\frac{\Delta+\ell+2\,\pi\,i\,T\,|k|}{2n}\right)\left(1+\frac{\Delta+\ell-2\,\pi\,i\,T\,|k|}{2n}\right)\right]^{(\ell+1)^2}=\\
&\prod_{k\in\mathbb{Z}}\prod_{\ell=0}^{\infty}\left[\frac{2\,\pi\,i\,T}{2}\left(|k|+\frac{\Delta+\ell}{2\,\pi\,i\,T}\right)\frac{-2\,\pi\,i\,T}{2}\left(|k|-\frac{\Delta+\ell}{2\,\pi\,i\,T}\right)\prod_{n=1}^{\infty}\frac{2\,\pi\,i\,T}{2}\left(|k|+\frac{2n+\Delta+\ell}{2\,\pi\,i\,T}\right)\frac{-2\,\pi\,i\,T}{2}\left(|k|-\frac{2n+\Delta+\ell}{2\,\pi\,i\,T}\right)\right]^{(\ell+1)^2}\\
\sim&\,\prod_{k\in\mathbb{Z}}\prod_{\ell=0}^{\infty}\left[\prod_{n=0}^{\infty}\left(|k|-i\,\frac{2n+\Delta+\ell}{2\,\pi\,T}\right)\left(|k|+i\,\frac{2n+\Delta+\ell}{2\,\pi\,T}\right)\right]^{(\ell+1)^2}
\end{aligned}
\end{equation}
where we used that the degeneration for each $\ell\ge 0$ is given by $(\ell+1)^2$. This again coincides with formula (2.10) in \cite{Law:2022zdq} using \eqref{AdSnormalmodes}.

 For the cases considered in \eqref{fulldetpuredS4} and \eqref{fulldetpureAdS5} we can also give the explicit formula for the $\zeta$-function regularized one-loop action as in \eqref{zetabh}. We have
\begin{equation}
\zeta_{\mathrm{dS_4}}(s) = \frac{1}{\Gamma(s)} \int_0^\infty\frac{dt}{t}\, t^s\, \frac{1+e^{-\beta t}}{1-e^{-\beta t}}
\sum_{\ell=0}^{\infty}\sum_{n=0}^{\infty} (2\ell+1)\,e^{i(\ell+n+1)t}
\end{equation}
in the four-dimensional de Sitter case, and
\begin{equation}
\zeta_{\mathrm{AdS_5}}(s) = \frac{1}{\Gamma(s)} \int_0^\infty\frac{dt}{t}\, t^s\, \frac{1+e^{-\beta t}}{1-e^{-\beta t}}
\sum_{\ell=0}^{\infty}\sum_{n=0}^{\infty} (\ell+1)^2\left[e^{(2\ell+n+\Delta)t}+e^{-(2\ell+n+\Delta)t} \right]
\end{equation}
in the five-dimensional anti-de Sitter case.

In the black hole problems, extracting explicitly the relevant factors as in \eqref{fulldetpureAdS5} from the Heun connection coefficients is complicated, 
since the zeros come from requiring the sum of the two channels in \eqref{finalresult} to vanish, and it is no longer possible to look only in the infinite product structure of the Gamma functions.

However, if we consider the small $R_h$ limit and the decomposition \eqref{conncoeffdecomposition}, the leading contribution of the determinant is given by
\begin{equation}
\begin{aligned}
\mathrm{det}\left(\Box-\mu^2\right)\sim\prod_{k\in\mathbb{Z}}\prod_{\ell,\vec{m}}\frac{2\pi\Gamma(2a)\Gamma(1+2a)t^{- a_0-a}}{\prod_{\sigma=\pm}\Gamma\left(\frac{1}{2}- a_0+a+\sigma\,a_t\right)\Gamma\left(\frac{1}{2}+a+ a_1+\sigma\,a_{\infty}\right)} \exp(-\frac{1}{2}\partial_{a_0}F(t)+\frac{1}{2}\partial_{a_1}F(t)+\frac{1}{2}\partial_aF(t)),
\end{aligned}
\end{equation}
where again the substitution $i\,\omega=2\,\pi\,T\,k$ is implied.

In all the considered cases, the leading order of $a$ in the expansion in $R_h$ (and also in the small $a_{\mathrm{BH}}$ expansion for the Kerr-de Sitter case) depends only on the angular quantum number $\ell$, and both indices $a_0$ and $a_t$ start with higher orders in $R_h$, therefore, neglecting the corrections in the last line in \eqref{conncoeffdecomposition}, the factors
\begin{equation}
\frac{2\pi\Gamma(2a)\Gamma(1+2a)t^{- a_0-a}}{\prod_{\sigma=\pm}\Gamma\left(\frac{1}{2}- a_0+a+\sigma\,a_t\right)} \exp(-\frac{1}{2}\partial_{a_0}F(t)+\frac{1}{2}\partial_{a_1}F(t)+\frac{1}{2}\partial_aF(t))
\end{equation}
only contribute as entire functions and do not give any contributions to zeros or poles in $\omega$, and all the analytic structure can be written by the infinite products in the Gamma functions 
\begin{equation}\label{gammafunctionpoleQNMs}
\Gamma\left(\frac{1}{2}+a+ a_1\pm a_{\infty}\right)\propto\frac{1}{\frac{1}{2}+a+ a_1\pm a_{\infty}}\prod_{n=1}^{\infty}\left(1+\frac{\frac{1}{2}+a+ a_1\pm a_{\infty}}{n}\right)^{-1}.
\end{equation}
Moreover, since $t\sim R_h$, only the leading order of $a$ contributes, and the reasoning proceeds as in the Hypergeometric cases.

In the final analytic structure, there is also one important difference between the Kerr-de Sitter case and the spherically symmetric cases, which is given by the degeneracies coming from the angular problem. In the spherically symmetric problems in four dimensions, these are given by $N^{(4)}(\ell)=2\ell+1$, and in five dimensions by $N^{(5)}(\ell)=(\ell+1)^2$, for each $\ell\ge 0$. 
In this approximation, the analytic structure in the leading order can therefore be read from
\begin{equation}\label{1}
\mathrm{det}\left(\Box-\mu^2\right)\sim\prod_{k\in\mathbb{Z}}\prod_{\ell=0}^{\infty}\left[\prod_{\sigma=\pm}\left(\frac{1}{2}+a+ a_1+\sigma\, a_{\infty}\right)\prod_{n=1}^{\infty}\prod_{\sigma=\pm}\left(1+\frac{\frac{1}{2}+a+ a_1+\sigma\, a_{\infty}}{n}\right)\right]^{N^{(d)}(\ell)},
\end{equation}
with $d=4,5$. In the Kerr-de Sitter case, instead, the formula reads
\begin{equation}\label{2}
\mathrm{det}\left(\Box-\mu^2\right)\sim\prod_{k\in\mathbb{Z}}\prod_{\ell=0}^{\infty}\prod_{m=-\ell}^{\ell}\left[\prod_{\sigma=\pm}\left(\frac{1}{2}+a+ a_1+\sigma\, a_{\infty}\right)\prod_{n=1}^{\infty}\prod_{\sigma=\pm}\left(1+\frac{\frac{1}{2}+a+ a_1+\sigma\, a_{\infty}}{n}\right)\right],
\end{equation}
and each pair of values $(\ell,m)$ gives a different contribution.

Although there are no closed expressions for the QNMs of the generic BH, 
we still can write approximate formulae by expanding in the BH radius $R_h$
by using the explicit power expansion of the QNMs \eqref{QNMexpansion}.
As these, to the first order can be found from the zeros of \eqref{1} and \eqref{2}, we get the following approximated expressions.

For the four-dimensional Schwarzschild-de Sitter case, one gets
\footnote{The Kerr-de Sitter case is more delicate, since the DHS formula applies to static spacetimes (for a generalization to the rotating cases see \cite{Castro:2017mfj,Arnaudo:2024bbd}).}
(see \cite{Aminov:2023jve})
\begin{equation}
\zeta_{\mathrm{SdS_4}}(s) = \frac{1}{\Gamma(s)} \int_0^\infty\frac{dt}{t}\, t^s\, \frac{1+e^{-\beta t}}{1-e^{-\beta t}}
\sum_{\ell=0}^{\infty}\sum_{n=0}^{\infty} (2\ell+1)\,e^{\left[i(\ell+n+1)-\omega_2\,R_h^2+\mathcal{O}(R_h^3)\right]t},
\end{equation}
where, for $\ell\ge 0$\footnote{For $\ell>0$ the correction $\omega_2$ can be found from the zeros of \eqref{1}.
 See equation $(3.23)$ in \cite{Aminov:2023jve} at $s=0$.
The case $\ell=0$ is more subtle. The full quantization condition \eqref{quantBperiod} must be used in this case. However, in the expansion in $R_h$, the leading order of the v.e.v. parameter $a$ equals $1/2$, and the NS function $F(t)$ has a pole for this value of the parameter. To find the analytic expression for $\omega_2$, we first assume $\ell$ to be generic in \eqref{quantBperiod}, and only in the final expansion in $R_h$ we send $\ell\to 0$.}
and $n\ge 0$,
\begin{equation}\label{omega2SdS}
\begin{aligned}
\omega_2=-\frac{i\left[\ell^3 \left(60 n^2+60 n+22\right)+\ell^2 \left(120 n^2+122n+45\right)+\ell \left(16 n^2 +19n+8\right)- \left(44 n^2+43n+15\right)\right]}{8 (\ell+1) (2 \ell+1)(2\ell-1)(2\ell+3)}.
\end{aligned}
\end{equation}

For the five-dimensional Schwarzschild-anti-de Sitter case one has, instead,
\begin{equation}
\begin{aligned}
\zeta_{\mathrm{SAdS_5}}(s) = \frac{1}{\Gamma(s)} \int_0^\infty\frac{dt}{t}\, t^s\, \frac{1+e^{-\beta t}}{1-e^{-\beta t}}
\sum_{\ell=0}^{\infty}\sum_{n=0}^{\infty} (\ell+1)^2\left\{e^{\left[2\ell+n+\Delta-\hat\omega_2\,R_h^2 +
\mathcal{O}(R_h^3)\right]t}+e^{-\left[2\ell+n+\Delta+ 
\hat\omega_2\,R_h^2 + \mathcal{O}(R_h^3)\right]t} \right\},
\end{aligned}
\end{equation}
where (see eq.(47) in \cite{Dodelson:2022yvn})
\begin{equation}
\hat\omega_2=-\frac{\Delta^2+\Delta(6n-1)+6n(n-1)}{2(\ell+1)}.
\end{equation}

The full result, rewritten in a form that makes explicit the analytic structure depending on the QNMs (see the quantization condition \eqref{quantBperiod}), is (for the equality in the formula see the comment in footnote \ref{footnoteprod})
\begin{equation}\label{3}
\begin{aligned}
\mathrm{det}\left(\Box-\mu^2\right)=&\,\prod_{k\in\mathbb{Z}}\prod_{\ell,\vec{m}}\Biggl\{\frac{2\pi\Gamma(2a)\Gamma(1+2a)}{\prod_{\sigma=\pm}\Gamma\left(\frac{1}{2}- a_0+a+\sigma\,a_t\right)\Gamma\left(\frac{1}{2}+a+ a_1+\sigma\,a_{\infty}\right)} t^{- a_0-a}\exp(-\frac{1}{2}\partial_{a_0}F(t)+\frac{1}{2}\partial_{a_1}F(t)+\frac{1}{2}\partial_aF(t))\times\\
&\,\times\left[1-\frac{\Gamma(-2a)^2\prod_{\sigma=\pm}\Gamma\left(\frac{1}{2}- a_0+a+\sigma\,a_t\right)\Gamma\left(\frac{1}{2}+a+ a_1+\sigma\,a_{\infty}\right)}{\Gamma(2a)^2\prod_{\sigma=\pm}\Gamma\left(\frac{1}{2}- a_0-a+\sigma\,a_t\right)\Gamma\left(\frac{1}{2}-a+ a_1+\sigma\,a_{\infty}\right)} t^{2a}\exp(-\partial_aF(t))\right]\Biggr\},
\end{aligned}
\end{equation}
where the substitution \eqref{thermalfrequancies} is implied, and the structure depending on QNMs can be read from the second line.

We remark again that, given a fixed $\ell_0$, the coefficients of the QNMs expansion \eqref{QNMexpansion} up to order $2\ell_0+1$ (for the four-dimensional cases) or $2\ell_0+2$ (for the five-dimensional case) in $R_h$ can be determined by the poles in \eqref{gammafunctionpoleQNMs}, where the additional complication compared to the Hypergeometric cases comes from the fact that $a$ is expressed as an instanton expansion and $\omega$ (or, equivalently, $k$) appears in the coefficients of such expansion.

\appendix

\section{Gelfand-Yaglom theorem}\label{appendixGY}

\subsection{Gelfand-Yaglom theorem for regular differential operators}

Let us introduce the setting in which the standard Gelfand-Yaglom theorem applies. Let
\begin{equation}\label{normal form}
\mathcal{D}=\frac{\mathrm{d}^2}{\mathrm{d}z^2}+V(z)
\end{equation}
be a second-order differential operator defined on the interval $z=[0,1]$. Let us consider the eigenvalue problem
\begin{equation}
\mathcal{D}\psi_n=\lambda_n\psi_n,
\end{equation}
with $\psi_n$ satisfying the Dirichlet boundary conditions
\begin{equation}
\psi_n(0)=\psi_n(1)=0,
\end{equation}
where $\{\lambda_n\}_n$ is the set of eigenvalues of $\mathcal{D}$, which is required to be discrete, non-degenerate, and bounded from below. Suppose we can solve the associated problem
\begin{equation}
\mathcal{D}\,u_{\lambda}=\lambda\,u_{\lambda},
\end{equation}
with $u_{\lambda}$ satisfying the boundary conditions
\begin{equation}\label{standardbc}
u_{\lambda}(0)=0,\quad u'_{\lambda}(0)=1.
\end{equation}
We call this $u_{\lambda}(z)$ the \emph{normalized solution} of $\left(\mathcal{D}-\lambda\right)u=0$ at $z=0$.

Then, one has that $u_{\lambda}(1)$ is equal to zero if and only if $\lambda$ is an eigenvalue of the operator $\mathcal{D}$. Indeed, $u_{\lambda}(1)=0$ if and only if $u_{\lambda}$ satisfies both the Dirichlet boundary conditions at $z=0$ and at $z=1$, but then $u_{\lambda}$ coincides with one of the eigenfunctions of $\mathcal{D}$, that is $\lambda=\lambda_n$ for some $n$.

Let $\tilde{\mathcal{D}}$ be some reference differential operator and $\tilde{u}_{\lambda}$ the corresponding eigenfunction satisfying \eqref{standardbc}. $\tilde{\mathcal{D}}$ is obtained by $\mathcal{D}$ by considering a deformation of the potential $V(z)$. It holds
\begin{equation}
\frac{\mathrm{det}\left(\mathcal{D}-\lambda\right)}{\mathrm{det}\left(\tilde{\mathcal{D}}-\lambda\right)}=\frac{u_{\lambda}(1)}{\tilde{u}_{\lambda}(1)}.
\end{equation}
Indeed, seen as functions of $\lambda$, both the LHS and the RHS have zeros in the eigenvalues of $\mathcal{D}$ and poles in the eigenvalues of $\tilde{\mathcal{D}}$. Therefore, the two must coincide up to a constant. Moreover,
\begin{equation}
\lim_{\lambda\to \infty}\frac{\mathrm{det}\left(\mathcal{D}-\lambda\right)}{\mathrm{det}\left(\tilde{\mathcal{D}}-\lambda\right)}=1,
\end{equation}
assuming the deformation of the potential $V(z)$ to be bounded and not modifying the asymptotic of the spectrum.
Hence, we can conclude that 
\begin{equation}
\frac{\mathrm{det}\left(\mathcal{D}\right)}{\mathrm{det}\left(\tilde{\mathcal{D}}\right)}=\frac{\mathrm{det}\left(\mathcal{D}-\lambda\right)}{\mathrm{det}\left(\tilde{\mathcal{D}}-\lambda\right)}\bigg|_{\lambda=0}=\frac{u_{\lambda=0}(1)}{\tilde{u}_{\lambda=0}(1)}.
\end{equation}

\begin{remark}\label{remarkGY} As a consequence of the theorem, we conclude that the ratio of determinants of two differential operators only depends on the normalized solutions of the corresponding differential equation.
\end{remark}

We also comment on the fact that it is not restrictive to consider differential operators in the normal form \eqref{normal form}. Indeed, let us consider a second-order differential equation of the form
\begin{equation}
\left[a(y)\frac{\mathrm{d}^2}{\mathrm{d}y^2}+b(y)\frac{\mathrm{d}}{\mathrm{d}y}+c(y)\right]\phi(y)=0,
\end{equation}
with the properties that $b(y)$ is differentiable and $a(y)$ is twice differentiable.
We can first redefine the variable as
\begin{equation}
\begin{aligned}
z=\frac{\int_0^y\mathrm{d}\bar{y}\frac{1}{\sqrt{a(\bar{y})}}}{C},\quad \text{with}\quad C=\int_0^1\mathrm{d}\bar{y}\frac{1}{\sqrt{a(\bar{y})}},
\end{aligned}
\end{equation}
so that the interval $y=[0,1]$ is mapped onto the interval $z=[0,1]$ and the differential equation becomes of the form
\begin{equation}
\left[\frac{\mathrm{d}^2}{\mathrm{d}z^2}+C\beta(z)\frac{\mathrm{d}}{\mathrm{d}z}+C^2\gamma(z)\right]\phi(z)=0,
\end{equation}
where
\begin{equation}
\beta(z)=\frac{b(y)-\frac{1}{2}a'(y)}{\sqrt{a(y)}},\quad \gamma(z)=c(y).
\end{equation}
Then, redefining the wave function $\phi$ as
\begin{equation}
\phi(z)=\exp(-\frac{1}{2}\int\mathrm{d}z\,C\beta(z))\psi(z),
\end{equation}
the differential equation becomes
\begin{equation}
\left[\frac{\mathrm{d}^2}{\mathrm{d}z^2}+V(z)\right]\psi(z)=0,
\end{equation}
with
\begin{equation}
V(z)=C^2\gamma(z)-\frac{C^2}{4}\beta(z)^2-\frac{C}{2}\beta'(z).
\end{equation}

\subsection{Gelfand-Yaglom version for regular singular points}

Suppose now that the potential $V(z)$ has regular singular points at $z=0$ and $z=1$. We denote with $\frac{1}{2}\pm a_0$ the roots of the indicial equation at $z=0$. Then, supposing not to be in a log case, there exists a fundamental system of solutions of the differential equation $\mathcal{D}\psi(z)=0$ around $z=0$ given by
\begin{equation}\label{localsolutionsat0}
\begin{aligned}
\psi_1^{(0)}=&\,z^{\frac{1}{2}+a_0}\left[1+\mathcal{O}(z)\right],\\
\psi_2^{(0)}=&\,z^{\frac{1}{2}-a_0}\left[1+\mathcal{O}(z)\right],
\end{aligned}
\end{equation}
and the Wronskian between the two solutions is (constant in $z$) equal to $2a_0$.

Let us suppose $\mathrm{Re}(a_0)>0$.
In order to apply the Gelfand-Yaglom theorem in the regular singular case, the standard vanishing Dirichlet boundary condition at $z=0$ is reformulated by asking the function $\psi$ to satisfy 
\begin{equation}
\lim_{z\to 0}\left(z^{\frac{1}{2}-a_0}\right)^{-1}\psi(z)=0.
\end{equation}
Analogous formulae hold at $z=1$.


Suppose that the points $z=0$ and $z=1$ are regular singular points of
both the equations $\mathcal{D}\psi(z)=0$ and $\tilde{\mathcal{D}}\tilde{\psi}(z)=0$ 
with equal indices $a_0=\tilde{a}_0$ and $a_1=\tilde{a}_1$. Suppose, moreover, $\mathrm{Re}(a_0)>0$ and $\mathrm{Re}(a_1)>0$.
Then,
\begin{equation}\label{gyF}
\frac{\mathrm{det}\left(\mathcal{D}\right)}{\mathrm{det}\left(\tilde{\mathcal{D}}\right)}=
\frac{\mathcal{C}_{12}}{\tilde{\mathcal{C}}_{12}}
\end{equation}
where $\mathcal{C}_{12}$ is the connection coefficient relating the local solutions around the two regular singular points:
\begin{equation}
\psi_1^{(0)}(z)=\mathcal{C}_{11}\psi_1^{(1)}(z)+\mathcal{C}_{12}\psi_2^{(1)}(z),
\end{equation}
and the same for $\tilde{\mathcal{C}}_{11}$ and $\tilde{\mathcal{C}}_{12}$.

Let us prove \eqref{gyF}. The strategy is to consider the associated problem as in the standard Gelfand-Yaglom theorem, but taking the normalized solution at a point close to $z=0$ in terms of the corresponding local solutions \eqref{localsolutionsat0}. Then, using the connection matrix, it is possible to analytically continue the solution close to the point $z=1$ and evaluate it there. Removing the cut-off, we get \eqref{gyF}.

The normalized solution at $z=0$ satisfying $u_{[\delta]}(\delta)=0$ and $u_{[\delta]}'(\delta)=1$
is given by 
\begin{equation}
u_{[\delta]}(z)=\frac{\psi_i^{(0)}(\delta)}{W(\psi_1^{(0)},\psi_2^{(0)})(\delta)}{\epsilon_{ij}} \psi_j^{(0)}(z)=\frac{\psi_i^{(0)}(\delta)}{2a_0}{\epsilon_{ij}} \psi_j^{(0)}(z),
\end{equation}
where 
\begin{equation}
\epsilon=\begin{pmatrix}
    0&1\\-1&0
\end{pmatrix}.
\end{equation}
Let us now analytically continue this solution to the neighborhood of $z=1$, as
\begin{equation}
\psi_i^{(0)}(z)=\mathcal{C}_{ij}\psi_j^{(1)}(z),
\end{equation}
and evaluate
\begin{equation}
u_{[\delta]}(1+\delta')=\frac{\psi_i^{(0)}(\delta)}{2a_0}{\epsilon_{ij}} \mathcal{C}_{jk}\psi_k^{(1)}(1+\delta')\, .
\end{equation}

By using the expansion of the local solutions around $z=0$ and $z=1$, and denoting
\begin{equation}
\begin{aligned}
&\rho_1^{(0)}=\frac{1}{2}+a_0,\quad \rho_2^{(0)}=\frac{1}{2}-a_0,\\
&\rho_1^{(1)}=\frac{1}{2}+a_1,\quad \rho_2^{(1)}=\frac{1}{2}-a_1,
\end{aligned}
\end{equation}
one finds
\begin{equation}
u_{[\delta]}(1+\delta')=
\frac{1}{2a_0}\left(\delta^{\rho_i^{(0)}}\right)
{\epsilon_{ij}} \mathcal{C}_{jk}\left(\delta'^{\rho_k^{(1)}}\right)
\left[1+\mathcal{O}\left(\delta,\delta'\right)\right].
\end{equation}

Consider now the ratio
\begin{equation}
\frac{u_{[\delta]}(1+\delta')}{\tilde{u}_{[\delta]}(1+\delta')},
\end{equation}
where the denominator is given by the above procedure for the reference operator $\tilde{\mathcal{D}}$ and with the same cut-off assignment.
As we remove the cut-off, in the limit $\delta,\delta'\to 0$ and using the assumptions $a_0,a_1>0$, the leading order term is given by
\begin{equation}
\frac{\mathrm{det}({\mathcal{D}})}{\mathrm{det}({\tilde{\mathcal{D}}})}=\lim_{\delta,\delta'\to0^+}
\frac{u_{[\delta]}(1 +\delta')}{\tilde{u}_{[\delta]}(1+\delta')}=
\lim_{\delta,\delta'\to0^+}
\frac{-\delta^{\frac{1}{2}-a_0}\mathcal{C}_{12}(\delta')^{\frac{1}{2}-a_1}}{-\delta^{\frac{1}{2}-a_0}\tilde{\mathcal{C}}_{12}(\delta')^{\frac{1}{2}-a_1}}= \frac{\mathcal{C}_{12}}{\tilde{\mathcal{C}}_{12}}.
\end{equation}

Similar results were obtained in \cite{lesch1999determinants}.

\section{Connection problems from \texorpdfstring{$z\sim 0$}{} to \texorpdfstring{$z\sim 1$}{}}\label{appendixA}

In this appendix, we recall the connection coefficients that analytically continue the local solutions around the singularity at $z=0$ in the region around the singularity at $z=1$ for the Heun and Hypergeometric differential operators.

\subsection{Hypergeometric connection formulae}

For the Hypergeometric equation
\begin{equation}\label{Hypergeometricdiffeq}
\left[z(1-z) \frac{d^2 }{dz^2}+\left(c-(a+b+1)z\right)\frac{d}{dz}-ab \right] w(z) = 0,
\end{equation}
a basis of local solutions around $z=0$ is given by
\begin{equation}
w_-^{(0)}(z)={}_2F_1(a,b;c;z),\quad w_+^{(0)}(z)=z^{1-c}{}_2F_1(a-c+1,b-c+1;2-c;z),
\end{equation}
and a basis of local solutions around $z=1$ is given by
\begin{equation}
w_-^{(1)}(z)={}_2F_1(a,b;a+b+1-c;1-z),\quad w_+^{(1)}(z)=(1-z)^{c-a-b}{}_2F_1(c-a,c-b;c-a-b+1;1-z).
\end{equation}
The connection formulae between these solutions are
\begin{equation}\label{hypergconnformulae}
\begin{aligned}
w_-^{(0)}(z)=&\,\frac{\Gamma(c)\Gamma(c-a-b)}{\Gamma(c-a)\Gamma(c-b)}w_-^{(1)}(z)+\frac{\Gamma(c)\Gamma(a+b-c)}{\Gamma(a)\Gamma(b)}w_+^{(1)}(z),\\ 
w_+^{(0)}(z)=&\,\frac{\Gamma(2-c)\Gamma(c-a-b)}{\Gamma(1-a)\Gamma(1-b)}w_-^{(1)}(z)+\frac{\Gamma(2-c)\Gamma(a+b-c)}{\Gamma(a-c+1)\Gamma(b-c+1)}w_+^{(1)}(z).
\end{aligned}
\end{equation}
Let us consider the normal form of the equation but supposing the index of the singularity at $z=\infty$ to satisfy $a_{\infty}^2=1/4$ (as we do in the normalizations introduced to the main part of the work):
\begin{equation}
\psi''(z) + \left[\frac{\frac{1}{4}-a_0^2}{z^2}+\frac{\frac{1}{4}-a_1^2}{(z-1)^2} - \frac{\frac{1}{2}-a_0^2-a_1^2 }{z(z-1)} \right]\psi(z)= 0,
\end{equation}
where the dictionary with the $a,b,c$ parameters is
\begin{equation}
a_0=\frac{1-c}{2},\quad a_1=\frac{c-a-b}{2},
\end{equation}
and with inverse
\begin{equation}
a=1-a_0-a_1,\quad b=-a_0-a_1,\quad c=1-2a_0.
\end{equation}
A basis of local solutions around $z=0$ is given by
\begin{equation}
\begin{aligned}
\psi_-^{(0)}(z)=&\,(z-1)^{\frac{1}{2}-a_1} z^{\frac{1}{2}-a_0}w_-^{(0)}(z)=\,(z-1)^{\frac{1}{2}+a_1}\,z^{\frac{1}{2}-a_0} \, _2F_1\left(-a_0+a_1,-a_0+a_1+1;1-2 a_0;z\right),\\
\psi_+^{(0)}(z)=&\,(z-1)^{\frac{1}{2}-a_1} z^{\frac{1}{2}-a_0}w_+^{(0)}(z)=\,(z-1)^{\frac{1}{2}+a_1}\,z^{\frac{1}{2}+a_0} \, _2F_1\left(a_0+a_1,a_0+a_1+1;1+2 a_0;z\right),
 \end{aligned}
\end{equation}
whose Wronskian is equal to $2a_0$, and a basis of local solutions around $z=1$ is given by
\begin{equation}
\begin{aligned}
\psi_-^{(1)}(z)=&\,(z-1)^{\frac{1}{2}-a_1} z^{\frac{1}{2}-a_0}w_-^{(1)}(z)=\,(z-1)^{\frac{1}{2}-a_1} z^{\frac{1}{2}+a_0}\, _2F_1\left(a_0-a_1,1+a_0-a_1;1-2 a_1;1-z\right),\\
\psi_+^{(1)}(z)=&\,(z-1)^{\frac{1}{2}-a_1} z^{\frac{1}{2}-a_0}w_+^{(1)}(z)=\,(z-1)^{\frac{1}{2}+a_1} z^{\frac{1}{2}+a_0}\, _2F_1\left(a_0+a_1,1+a_0+a_1;1+2 a_1;1-z\right),
\end{aligned}
\end{equation}
whose Wronskian is equal to $2a_1$.
The corresponding connection formulae read
\begin{equation}
\begin{aligned}
\psi_-^{(0)}(z)=&\,\frac{\Gamma(1-2a_0)\Gamma(2a_1)}{\Gamma(-a_0+a_1)\Gamma(1-a_0+a_1)}\psi_-^{(1)}(z)+\frac{\Gamma(1-2a_0)\Gamma(-2a_1)}{\Gamma(1-a_0-a_1)\Gamma(-a_0-a_1)}\psi_+^{(1)}(z),\\
\psi_+^{(0)}(z)=&\,\frac{\Gamma(1+2a_0)\Gamma(2a_1)}{\Gamma(a_0+a_1)\Gamma(1+a_0+a_1)}\psi_-^{(1)}(z)+\frac{\Gamma(1+2a_0)\Gamma(-2a_1)}{\Gamma(1+a_0-a_1)\Gamma(a_0-a_1)}\psi_+^{(1)}(z).
\end{aligned}
\end{equation}

\subsection{Heun connection formula for \texorpdfstring{$|t|<1$}{}}

The analogous problem was solved for the Heun equation in \cite{Bonelli:2022ten}, where connection formulae between semiclassical Liouville conformal blocks were studied. In the main part of the work, we consider the regime in which $|t|<1$. In the next subsection, we show the analogous formulae in the regime $|t|>1$.

The conformal block for small $t$ around $z=0$ reads
\begin{equation}
\FIV{\alpha_\infty}{\alpha_1}{\alpha}{\alpha_t}{{\alpha_{0 \theta}}}{\alpha_{2,1}}{\alpha_0}{t}{\frac{z}{t}} \,.
\end{equation}
The conformal block for small $t$ around $z=1$ reads
\begin{equation}
t^{\Delta_\infty + \Delta_1 + \Delta_{2,1} - \Delta_t - \Delta_0} (1-t)^{\Delta_\infty + \Delta_0 + \Delta_{2,1} - \Delta_t - \Delta_1} (z-t)^{- 2 \Delta_{2,1}} \FIV{\alpha_t}{\alpha_0}{\alpha}{\alpha_\infty}{\alpha_{1 \theta}}{\alpha_{2,1}}{\alpha_1}{t}{\frac{z-1}{z-t}}.
\end{equation}
In the semiclassical limit, these read
\begin{equation}
\begin{aligned}
&\FIVsc{a_\infty}{a_1}{a}{a_t}{{a_{0 \theta}}}{a_{2,1}}{a_0}{t}{\frac{z}{t}},\\
&\left( t(1-t) \right)^{-\frac{1}{2}} (z-t) \FIVsc{a_t}{a_0}{a}{a_\infty}{{a_{1 \theta}}}{a_{2,1}}{a_1}{t}{\frac{z-1}{z-t}}.
\end{aligned}
\end{equation}
The connection formula between the two semiclassical blocks, written in terms of the connection matrices of the Hypergeometric functions
\begin{equation}
\mathcal{M}_{\theta \theta'}(a_1,a_2;a_3) = \frac{\Gamma(-2\theta'a_2)\Gamma(1+2\theta a_1)}{\Gamma\left(\frac{1}{2}+\theta a_1-\theta' a_2 + a_3\right) \Gamma\left(\frac{1}{2}+\theta a_1-\theta' a_2 - a_3\right)},\ \ \text{where\ }\ \theta,\theta'=\pm
\end{equation}
reads
\begin{equation}
\begin{aligned}
\FIVsc{a_\infty}{a_1}{a}{a_t}{{a_{0 \theta}}}{a_{2,1}}{a_0}{t}{\frac{z}{t}}=&\,\sum_{\theta',\theta''=\pm}\mathcal{M}_{\theta\theta'}(a_0,a;a_t)\mathcal{M}_{(-\theta')\theta''}(a,a_1;a_{\infty})t^{\theta' a}\exp(-\frac{\theta'}{2}\partial_aF(t))\times\\
&\times\left( t(1-t) \right)^{-\frac{1}{2}} (z-t) \FIVsc{a_t}{a_0}{a}{a_\infty}{{a_{1 \theta''}}}{a_{2,1}}{a_1}{t}{\frac{z-1}{z-t}},
\end{aligned}
\end{equation}
$F(t)$ being the classical 4-point conformal block (see Appendix~\ref{appendixNS})
\begin{equation}\label{classical4point}
F(t)=F\left(\begin{matrix} a_1\\ a_{\infty}
\end{matrix}\, a\,\begin{matrix} a_t\\ a_{0}
\end{matrix}\,;\,t\right).
\end{equation}
We also need to relate the expansions of these semiclassical blocks to the local solutions of the Heun equation in normal form
\begin{equation}\label{solutionnormalform}
\begin{aligned}
&\psi_{\theta}^{(0)}(z)\sim z^{\frac{1}{2}+\theta a_0}\left[1+\mathcal{O}(z)\right],\\
&\psi_{\theta}^{(1)}(z)\sim (z-1)^{\frac{1}{2}+\theta a_1}\left[1+\mathcal{O}(z-1)\right].
\end{aligned}
\end{equation}
The semiclassical blocks' expansions read
\begin{equation}
\begin{aligned}
&\FIVsc{a_\infty}{a_1}{a}{a_t}{{a_{0 \theta}}}{a_{2,1}}{a_0}{t}{\frac{z}{t}}\sim t^{-\theta a_0}\exp(-\frac{\theta}{2}\partial_{a_0}F(t))z^{\frac{1}{2}+\theta a_0}\left[1+\mathcal{O}\left(t,\frac{z}{t}\right)\right],\\
&\FIVsc{a_t}{a_0}{a}{a_\infty}{{a_{1 \theta''}}}{a_{2,1}}{a_1}{t}{\frac{z-1}{z-t}}\sim \left(\frac{t}{1-t}\right)^{1/2}\left(z-1\right)^{\frac{1}{2}+\theta'' a_1}\exp(-\frac{\theta''}{2}\partial_{a_1}F(t))\left[1+\mathcal{O}\left(t,\frac{z-1}{z-t}\right)\right],
\end{aligned}
\end{equation}
where $F(t)$ is the conformal block \eqref{classical4point}.

It follows that the connection formula between the solutions of the Heun equation reads
\begin{equation}\label{connection0to1}
\begin{aligned}
&\psi_{\theta}^{(0)}(z)=\,\sum_{\theta''=\pm}\mathcal{C}_{\theta \theta''}\,\psi_{\theta''}^{(1)}(z),\quad\text{with}\\
&\mathcal{C}_{\theta \theta''}=\sum_{\theta'=\pm}\mathcal{M}_{\theta\theta'}(a_0,a;a_t)\mathcal{M}_{(-\theta')\theta''}(a,a_1;a_{\infty})t^{\theta a_0+\theta' a}\exp(\frac{\theta}{2}\partial_{a_0}F(t)-\frac{\theta''}{2}\partial_{a_1}F(t)-\frac{\theta'}{2}\partial_aF(t)).
\end{aligned}
\end{equation}

\subsection{Heun connection formula for \texorpdfstring{$|t|>1$}{}}\label{Heunconnectiont>1}

The connection formula between the points $z=0$ and $z=1$ in the regime $|t|>1$ is simpler since the two singular points are on the same side of the pants decomposition of the four-punctured sphere.

In the semiclassical limit, the conformal blocks around $z=0$ and $z=1$ read
\begin{equation}
\begin{aligned}
&t^{1/2}\mathcal{F}\left(\begin{matrix}a_t\\ a_{\infty}\end{matrix}\, a\, \begin{matrix}a_1\\ {}\end{matrix}\, a_{0\theta}\,\begin{matrix}a_{2,1}\\ a_0\end{matrix};\frac{1}{t},z\right),\\
&(t-1)^{1/2}e^{\theta i \pi a}\mathcal{F}\left(\begin{matrix}a_t\\ a_{\infty}\end{matrix}\, a\, \begin{matrix}a_0\\ {}\end{matrix}\, a_{1\theta}\,\begin{matrix}a_{2,1}\\ a_1\end{matrix};\frac{1}{t-1},1-z\right),
\end{aligned}
\end{equation}
respectively. The connection formula between them reads
\begin{equation}
\begin{aligned}
t^{\frac{1}{2}} \FIVsc{a_\infty}{a_t}{a}{a_1}{{a_{0 \theta}}}{a_{2,1}}{a_0}{\frac{1}{t}}{z} = \sum_{\theta' = \pm} \mathcal{M}_{\theta \theta'} \left( a_0, a_1 ; a \right) (t-1)^{\frac{1}{2}} e^{\theta' i \pi a} \FIVsc{a_\infty}{a_t}{a}{a_1}{{a_{0 \theta'}}}{a_{2,1}}{a_0}{\frac{1}{t-1}}{1-z} \,.
\end{aligned}
\end{equation}
The semiclassical blocks' expansions read
\begin{equation}
\begin{aligned}
&\mathcal{F}\left(\begin{matrix}a_t\\ a_{\infty}\end{matrix}\, a\, \begin{matrix}a_1\\ {}\end{matrix}\, a_{0\theta}\,\begin{matrix}a_{2,1}\\ a_0\end{matrix};\frac{1}{t},z\right)\sim t^{-\frac{1}{2}}\exp(-\frac{\theta}{2}\partial_{a_0}F(t))z^{\frac{1}{2}+\theta a_0}\left[1+\mathcal{O}\left(\frac{1}{t},z\right)\right],\\
&\FIVsc{a_\infty}{a_t}{a}{a_0}{{a_{1 \theta'}}}{a_{2,1}}{a_1}{\frac{1}{t-1}}{1-z}\sim (t-1)^{-\frac{1}{2}}\,e^{\theta'\,i\,\pi\,a}\,\left(z-1\right)^{\frac{1}{2}+\theta' a_1}\exp(-\frac{\theta'}{2}\partial_{a_1}F(t))\left[1+\mathcal{O}\left(\frac{1}{t-1},1-z\right)\right],
\end{aligned}
\end{equation}
where $F(t)$ is the conformal block 
\begin{equation}
F=F\left(\begin{matrix} a_t\\ a_{\infty}\end{matrix}\, a\, \begin{matrix} a_1\\ a_{0}\end{matrix};\frac{1}{t}\right).
\end{equation}

It follows that the connection formula between the solutions of the Heun equation with the expansion \eqref{solutionnormalform} reads
\begin{equation}
\begin{aligned}
&\psi_{\theta}^{(0)}(z)=\,\sum_{\theta'=\pm}\mathcal{C}_{\theta \theta'}\,\psi_{\theta'}^{(1)}(z),\quad\text{with}\\
&\mathcal{C}_{\theta \theta'}=\mathcal{M}_{\theta\theta'}(a_0,a_1;a)\exp(\frac{\theta}{2}\partial_{a_0}F(t)-\frac{\theta'}{2}\partial_{a_1}F(t))
\end{aligned}
\end{equation}

\section{Determinant of Heun differential operators and Gelfand-Yaglom theorem}\label{appendixHeunoverhyperg}

\subsection{Heun normalized with Hypergeometric}\label{ratiodeterminants}

In the main part of the work we need to compute (ratio of) determinants of differential operators of Heun's and Hypergeometric's type. 

Indeed, the spectral problems in which we are interested are encoded by Heun differential equations with both boundary conditions imposed at singular points. If we consider the normal form of the Heun operator
\begin{equation}\label{heundiffop}
\mathcal{D}:\quad\frac{\mathrm{d}^2}{\mathrm{d}z^2} + \left[\frac{\frac{1}{4}-a_0^2}{z^2}+\frac{\frac{1}{4}-a_1^2}{(z-1)^2} + \frac{\frac{1}{4}-a_t^2}{(z-t)^2}- \frac{\frac{1}{2}-a_1^2 -a_t^2 -a_0^2 +a_\infty^2 + u}{z(z-1)}+\frac{u}{z(z-t)} \right],
\end{equation}
the simpler problem can be taken to be a Hypergeometric operator, which can be obtained by modifying the potential setting $u=0$, $a_t^2=\frac{1}{4}$. For simplicity, we also set $a_{\infty}^2=\frac{1}{4}$. This gives
\begin{equation}\label{hypergdiffop}
\tilde{\mathcal{D}}:\quad\frac{\mathrm{d}^2}{\mathrm{d}z^2} + \left[\frac{\frac{1}{4}-a_0^2}{z^2}+\frac{\frac{1}{4}-a_1^2}{(z-1)^2}- \frac{\frac{1}{2}-a_1^2 -a_0^2}{z(z-1)} \right].
\end{equation}

This simplification is such that the indices of the singularities at $z=0$ and $z=1$ are kept fixed. Using the connection coefficients for the Heun equation and Hypergeometric equation (see Appendix~\ref{appendixA}), we can take the ratio of determinants as the ratio of connection coefficients, distinguishing the cases according to the signs of $\mathrm{Re}(a_0)$ and $\mathrm{Re}(a_1)$.

For example, in the case $\mathrm{Re}(a_0)>0$ and $\mathrm{Re}(a_1)>0$, we have
\begin{equation}\label{ratio1}
\begin{aligned}
\frac{\mathrm{det}\left(\mathcal{D}\right)}{\mathrm{det}\left(\tilde{\mathcal{D}}\right)}=\frac{\sum_{\theta'=\pm}\mathcal{M}_{+\theta'}(a_0,a;a_t)\mathcal{M}_{(-\theta')-}(a,a_1;a_{\infty})t^{a_0+\theta' a}\exp(\frac{1}{2}\partial_{a_0}F(t)+\frac{1}{2}\partial_{a_1}F(t)-\frac{\theta'}{2}\partial_aF(t))}{\frac{\Gamma(1+2a_0)\Gamma(2a_1)}{\Gamma(1+a_0+a_1)\Gamma(a_0+a_1)}}.
\end{aligned}
\end{equation}

\subsection{Computation of determinant for Hypergeometric operators}\label{determinantofhyperg}

In this Appendix, we use a different method to compute the determinant of generic Hypergeometric differential operators. Using the result of the ratio of determinants \eqref{ratio1}, this gives a prescription on how to compute the (regularized) determinant for Heun differential operators. 

Let $\mathcal{D}_1$ be the Hypergeometric differential operator in normal form with generic indices of the singularities, parametrized by the parameters $a,b,c$:
\begin{equation}\label{hypergeodiffop}
\mathcal{D}_1:\quad \frac{\mathrm{d}^2}{\mathrm{d}z^2}+\frac{2 c [z (a+b-1)+1]+z \left[-z (a-b)^2-4 a b+z\right]-c^2}{4 (z-1)^2 z^2}.
\end{equation}

Let $\mathcal{D}_2$ be the Hypergeometric differential operator in the form
\begin{equation}
\mathcal{D}_2:\quad \frac{\mathrm{d}^2}{\mathrm{d}z^2}+\frac{\left[c-(a+b+1)z\right]}{z(1-z)}\frac{\mathrm{d}}{\mathrm{d}z}-\frac{a\,b}{z(1-z)}.
\end{equation}
We have that if $\psi_{1,\lambda}(z)$ is an eigenfunction for $\mathcal{D}_1$ with corresponding eigenvalue $\lambda$, then
\begin{equation}
\psi_{2,\lambda}(z)=z^{-c/2} (z-1)^{-\frac{a+b+1-c}{2}}\psi_{1,\lambda}(z)
\end{equation}
is an eigenfunction for $\mathcal{D}_2$ with the same eigenvalue $\lambda$. Indeed,
\begin{equation}
\begin{aligned}
\mathcal{D}_2\psi_{2,\lambda}(z)=&\,\left[z^{-c/2}(z-1)^{-(a+b+1-c)/2}\mathcal{D}_1\,z^{c/2}(z-1)^{(a+b+1-c)/2}\right]\left[z^{-c/2}(z-1)^{-(a+b+1-c)/2}\psi_{1,\lambda}(z)\right]=\\
=&\,z^{-c/2}(z-1)^{-(a+b+1-c)/2}\mathcal{D}_1\psi_{1,\lambda}(z)=z^{-c/2}(z-1)^{-(a+b+1-c)/2}\lambda\psi_{1,\lambda}(z)=\lambda\psi_{2,\lambda}(z).
\end{aligned}
\end{equation}
Therefore, the determinant of the two differential operators is the same, since the two have the same eigenvalues.

Now, thanks to the Gelfand-Yaglom theorem and the remark \ref{remarkGY}, the determinant of $\mathcal{D}_2$ is equal to the determinant of the operator
\begin{equation}
\mathcal{D}_3:\quad z(1-z)\frac{\mathrm{d}^2}{\mathrm{d}z^2}+\left[c-(a+b+1)z\right]\frac{\mathrm{d}}{\mathrm{d}z}-a\,b,
\end{equation}
since the differential equations $\mathcal{D}_2\psi(z)=0$ and $\mathcal{D}_3\psi(z)=0$ have the same solutions. Indeed, to apply the Gelfand-Yaglom theorem and compute the determinants of $\mathcal{D}_2$ and $\mathcal{D}_3$, one can transform them in the normal form using the procedure outlined in Appendix~\ref{appendixGY}, and finally, normalizing with respect to the same reference operator, one finds the same result for the two computations.

Finally, in order to compute the determinant of $\mathcal{D}_3$, we can look into the eigenvalue problem
\begin{equation}
\left(\mathcal{D}_3-\lambda\right)w(z)=0.
\end{equation}
A basis of independent solutions of this differential equation around $z=0$ is given by
\begin{equation}
\begin{aligned}
&w_-^{(0)}(z)=\, _2F_1\left(-\frac{1}{2} \sqrt{a^2-2 a b+b^2-4 \lambda }+\frac{a}{2}+\frac{b}{2},\frac{1}{2} \sqrt{a^2-2 a b+b^2-4 \lambda }+\frac{a}{2}+\frac{b}{2};c;z\right),\\
&w_+^{(0)}(z)=z^{1-c}\, _2F_1\left(-\frac{1}{2} \sqrt{a^2-2 a b+b^2-4 \lambda }+\frac{a}{2}+\frac{b}{2}-c+1,\frac{1}{2} \sqrt{a^2-2 a b+b^2-4 \lambda }+\frac{a}{2}+\frac{b}{2}-c+1;2-c;z\right).
\end{aligned}
\end{equation}
The selected solution around $z=0$ is $w_+^{(0)}(z)$. The connection coefficient in front of the discarded solution around $z=1$ is given by
\begin{equation}
\frac{\Gamma\left(2-c\right)\Gamma\left(c-a-b\right)}{\Gamma\left(1+\frac{1}{2} \sqrt{a^2-2 a b+b^2-4 \lambda }-\frac{a}{2}-\frac{b}{2}\right)\Gamma\left(1-\frac{1}{2} \sqrt{a^2-2 a b+b^2-4 \lambda }-\frac{a}{2}-\frac{b}{2}\right)}.
\end{equation}
Therefore, the $\lambda_n$ that ensure the correct boundary conditions for the solution are obtained by the quantization condition
\begin{equation}
1+\frac{1}{2} \sqrt{a^2-2 a b+b^2-4 \lambda_n }-\frac{a}{2}-\frac{b}{2}=-n\quad\text{or}\quad 1-\frac{1}{2} \sqrt{a^2-2 a b+b^2-4 \lambda_n }-\frac{a}{2}-\frac{b}{2}=-n,\ \ n\in\mathbb{Z}_{\ge 0},
\end{equation}
that is,
\begin{equation}
\lambda_n=(1-b+n)(-1+a-n).
\end{equation}
Hence, denoting with a tilde the regularization of the previous infinite product, the determinant is given by
\begin{equation}
\mathrm{det}\mathcal{D}_3=\widetilde{\prod}_{n\ge 0}(1-b+n)(-1+a-n)=\frac{2\pi}{\Gamma(1-b)\Gamma(1-a)},
\end{equation}
where we used the Zeta regularization and the \emph{Lerch's formula} \cite{iwasawa}.

\subsection{Regularized determinant for Heun differential operators}\label{appendixHeundeterminant}

Comparing the differential operator $\mathcal{D}_1$ in \eqref{hypergeodiffop} with the operator $\tilde{\mathcal{D}}$ in \eqref{hypergdiffop}, we have that the two are related by the dictionary\footnote{In \eqref{C.15} the parameter $a$ denotes the parameter of the Hypergeometric equation, whereas in \eqref{C.17} it denotes the v.e.v. parameter.}
\begin{equation}\label{C.15}
a=1-a_0-a_1,\quad\text{and}\quad b=-a_0-a_1.
\end{equation}
With the result of the previous subsection, we have that
\begin{equation}
\mathrm{det}\tilde{\mathcal{D}}=\widetilde{\prod}_{n\ge 0}(1+a_0+a_1+n)(-a_0-a_1-n)=\frac{2\pi}{\Gamma(1+a_0+a_1)\Gamma(a_0+a_1)}.
\end{equation}
We conclude that we can give a formula for the (regularized) determinant of the Heun differential operator \eqref{heundiffop}, under the assumption $\mathrm{Re}(a_0)>0$ and $\mathrm{Re}(a_1)>0$:
\begin{equation}\label{C.17}
\begin{aligned}
\mathrm{det}\mathcal{D}=&\sum_{\theta'=\pm}\frac{2\pi\Gamma(-2\theta'a)\Gamma(1-2\theta'a)}{\prod_{\sigma=\pm}\Gamma\left(\frac{1}{2}+a_0-\theta'a+\sigma\,a_t\right)\Gamma\left(\frac{1}{2}-\theta'a+a_1+\sigma\,a_{\infty}\right)}t^{a_0+\theta' a}\exp(\frac{1}{2}\partial_{a_0}F(t)+\frac{1}{2}\partial_{a_1}F(t)-\frac{\theta'}{2}\partial_aF(t)).
\end{aligned}
\end{equation}
Let us remark that this result is equal to the Heun connection coefficient in front of the discarded solution at $z=1$, divided by the two Gamma functions whose arguments depend on the indices of the singularities where the two boundary conditions are imposed. The $2\pi$ factor comes from the Zeta function regularization. 
This normalization gives analogous results as the ones obtained in the work \cite{Law:2022zdq}, where the subtraction of the Gamma functions is motivated by physical arguments, introducing the \emph{Rindler-like region}.

\subsection{Example of equality of determinants}

We show explicitly that the ratio of the determinants of the differential operators where one is obtained by multiplying the other for $z(1-z)$ is equal to 1 in the simplest case, in which we take 
\begin{equation}
\mathcal{D}_1=-\frac{\mathrm{d^2}}{\mathrm{d}z^2},\quad \mathcal{D}_2=-z(1-z)\frac{\mathrm{d^2}}{\mathrm{d}z^2}.
\end{equation}
This is just a consistency check of the previous remark that comes from the proof of the standard version of the Gelfand-Yaglom theorem.
For the determinants of these operators, we can use the standard form of the Gelfand-Yaglom theorem, and consider the associated problems
\begin{equation}
\begin{aligned}
&\mathcal{D}_1u_{1,\lambda}=\lambda u_{1,\lambda},\quad \text{with}\ \ u_{1,\lambda}(0)=0\ \ \text{and}\ \ u'_{1,\lambda}(0)=1,\\
&\mathcal{D}_2u_{2,\lambda}=\lambda u_{2,\lambda},\quad \text{with}\ \ u_{2,\lambda}(0)=0\ \ \text{and}\ \ u'_{2,\lambda}(0)=1.
\end{aligned}
\end{equation}
The solutions satisfying these boundary conditions are given by
\begin{equation}
\begin{aligned}
&u_{1,\lambda}=\frac{\sin(\sqrt{\lambda} z)}{\sqrt{\lambda}},\\
&u_{2,\lambda}=z \, _2F_1\left(\frac{1}{2}-\frac{1}{2} \sqrt{4 \lambda +1},\frac{1}{2} \sqrt{4 \lambda +1}+\frac{1}{2};2;z\right).
\end{aligned}
\end{equation}
Therefore, the ratio of determinants of the original operators is given by
\begin{equation}
\frac{\mathrm{det}\left(\mathcal{D}_1\right)}{\mathrm{det}\left(\mathcal{D}_2\right)}=\frac{u_{1,\lambda=0}(1)}{u_{2,\lambda=0}(1)}=\frac{1}{1}=1.
\end{equation}
This could be seen more easily from the standard version of the Gelfand-Yaglom theorem, since in both cases the solution $\bar{\psi}(z)$ of the differential equation $\mathcal{D}_i\psi(z)=0$ satisfying $\psi(0)=0$ and $\psi'(0)=1$ is given by $\bar{\psi}(z)=z$, and we have $\bar{\psi}(1)=1$.

\section{Reduction of connection coefficient}

Here, we prove that in the limit in which the Heun differential operator $\mathcal{D}$ reduces to the Hypergeometric one, the ratio in \eqref{ratio1} becomes equal to 1. Notice that we are not taking a collision limit, but we are just fitting the parameters so that the singularity at $z=t$ becomes an apparent one.

The differential operator $\mathcal{D}$ in \eqref{heundiffop} reduces to $\tilde{\mathcal{D}}$ in \eqref{hypergdiffop} by setting
\begin{equation}\label{reduceddictio}
a_t=a_{\infty}=\frac{1}{2},\quad 
u=0.
\end{equation}
Using the instanton expansion
\begin{equation}
u=-\frac{1}{4}+a_t^2+a_0^2-a^2+t\frac{\partial F(t)}{\partial t},
\end{equation}
we get 
\begin{equation}
a=\pm a_0,\quad F(t)=0.
\end{equation}
Let us choose the plus sign.
In the Hypergeometric connection matrices appearing in the determinant of $\mathcal{D}$
\begin{equation}
\begin{aligned}
\mathcal{M}_{+\theta'}(a_0,a;a_t)\mathcal{M}_{(-\theta')-}(a,a_1;a_{\infty})=\frac{\Gamma\left(1+2a_0\right)\Gamma\left(-2\theta'a\right)\Gamma\left(1-2\theta'a\right)\Gamma\left(2a_1\right)}{\Gamma\left(\frac{1}{2}+a_0-\theta'a+a_t\right)\Gamma\left(\frac{1}{2}+a_0-\theta'a-a_t\right)\Gamma\left(\frac{1}{2}-\theta'a+a_1+a_{\infty}\right)\Gamma\left(\frac{1}{2}-\theta'a+a_1-a_{\infty}\right)},
\end{aligned}
\end{equation}
one can see that the choice $\theta'=+$ makes one of the arguments of the Gamma functions in the denominator equal to 0 under the dictionary \eqref{reduceddictio} (if we had chosen the different sign $a_0=-a$, then the reasoning would have been the same with $\theta'=-$). Therefore, the determinant of $\mathcal{D}$ reduces to channel corresponding to $\theta'=-$:
\begin{equation}
\begin{aligned}
&\frac{\Gamma\left(1+2a_0\right)\Gamma\left(2a\right)\Gamma\left(1+2a\right)\Gamma\left(2a_1\right)}{\Gamma\left(\frac{1}{2}+a_0+a+a_t\right)\Gamma\left(\frac{1}{2}+a_0+a-a_t\right)\Gamma\left(\frac{1}{2}+a+a_1+a_{\infty}\right)\Gamma\left(\frac{1}{2}+a+a_1-a_{\infty}\right)}t^{a_0-a}\exp(\frac{1}{2}\partial_{a_0}F(t)+\frac{1}{2}\partial_{a_1}F(t)+\frac{1}{2}\partial_aF(t))\to\\
&\to \frac{\Gamma\left(1+2a_0\right)^2\Gamma\left(2a_0\right)\Gamma\left(2a_1\right)}{\Gamma\left(1+2a_0\right)\Gamma\left(2a_0\right)\Gamma\left(1+a_0+a_1\right)\Gamma\left(a_0+a_1\right)}=\frac{\Gamma\left(1+2a_0\right)\Gamma\left(2a_1\right)}{\Gamma\left(1+a_0+a_1\right)\Gamma\left(a_0+a_1\right)}
\end{aligned}
\end{equation}
where we used \eqref{reduceddictio} to pass to the second line. The last result is precisely the determinant of $\tilde{\mathcal{D}}$ appearing in the denominator of \eqref{ratio1} (equivalently, one of the Hypergeometric connection coefficients).

\section{Gauge theory conventions}\label{appendixNS}

This appendix collects the notations and conventions used when applying the gauge theory approach to the Heun connection problems. The relevant theory is $\mathcal{N}=2$ $SU(2)$ gauge theory with $N_f=4$ fundamental hypermultiplets. 

If $Y$ is a Young diagram, we denote with $(Y_1\ge Y_2\ge\dots)$ the heights of its columns and with $(Y'_1\ge Y'_2,\dots)$ the lengths of its rows. For every Young diagram $Y$ and for every box $s=(i,j)$, we denote the arm length and the leg length of $s$ with respect to the diagram $Y$ as
\begin{equation}
A_Y(i, j) = Y_j -  i, \quad L_Y(i, j) =Y'_i - j.
\end{equation}

We now introduce the main contributions crucial for the definition of the instanton partition function of $\mathcal{N}=2$ $SU(2)$  gauge theory with fundamental matter. Let us denote with $\vec{Y}=\left( Y_1, Y_2 \right)$ a pair of Young diagrams. We denote with $\vec{a}=(a_1,a_2)$ the v.e.v. of the scalar in the vector multiplet and with $\epsilon_1,\epsilon_2$ the parameters characterizing the $\Omega$-background. We define the hypermultiplet and vector contribution as \cite{Flume:2002az,Bruzzo:2002xf}
\begin{equation}
\begin{aligned}
    &z_{\text{hyp}} \left( \vec{a}, \vec{Y}, m \right) = \prod_{k= 1,2} \prod_{(i,j) \in Y_k} \left[ a_k + m + \epsilon_1 \left( i - \frac{1}{2} \right) + \epsilon_2 \left( j - \frac{1}{2} \right) \right] \,, \\
    &z_{\text{vec}} \left( \vec{a}, \vec{Y} \right) = \prod_{i,j=1}^2\prod_{s\in Y_i}\frac{1}{a_i-a_j-\epsilon_1L_{Y_j}(s)+\epsilon_2(A_{Y_i}(s)+1)}\prod_{t\in Y_j}\frac{1}{-a_j+a_i+\epsilon_1(L_{Y_i}(t)+1)-\epsilon_2A_{Y_j}(s)}\,.
\end{aligned}
\end{equation}
We always take $\epsilon_1=1$ and $\vec{a}=(a,-a)$.
Let us denote with $m_1,m_2,m_3,m_4$ the masses of the four hypermultiplets and let us introduce the gauge parameters $a_0,a_t,a_1,a_{\infty}$ satisfying
\begin{equation}\label{gaugemasses}
\begin{aligned}
m_1&=-a_t-a_0,\quad
&m_2=-a_t+a_0,\\
m_3&=a_{\infty}+a_1,\quad
&m_4=-a_{\infty}+a_1.
\end{aligned}
\end{equation}
Moreover, we denote with $t$ the instanton counting parameter $t=e^{2\pi i\tau}$, where $\tau$ is related to the gauge coupling by 
\begin{equation}
    \tau=\frac{\theta}{2\pi}+i\frac{4\pi}{g_{\rm YM}^2}.
\end{equation}
The instanton part of the NS free energy is then given as a power series in $t$ by
\begin{equation}
F(t)=\lim_{\epsilon_2\to 0}\epsilon_2\log\Biggl[(1-t)^{-2\epsilon_2^{-1}\left(\frac{1}{2}+a_1\right)\left(\frac{1}{2}+a_t\right)}\sum_{\vec{Y}}t^{|\vec{Y}|}z_{\text{vec}} \left( \vec{a}, \vec{Y} \right)\prod_{i=1}^4z_{\text{hyp}} \left( \vec{a}, \vec{Y}, m_i \right)\Biggr].
\end{equation}

The gauge parameter $a$ is expressed in a series expansion in the instanton counting parameter $t$, obtained by inverting the \emph{Matone relation} \cite{Matone:1995rx}
\begin{equation}
u =-\frac{1}{4} - a^2 + a_t^2 + a_0^2 + t \partial_t F(t),
\end{equation}
where the parameter $u$ appearing in the differential equation \eqref{heunnormalform} is the complex moduli parametrizing the corresponding SW curve.
Explicitly, the expansion reads as follows
\begin{equation} 
a=\pm\left\{\sqrt{-\frac{1}{4}-u+a_t^2+a_0^2}+\frac{\bigl(\frac{1}{2}+u-a_t^2-a_0^2-a_1^2+a_{\infty}^2\Bigr)\Bigl(\frac{1}{2}+u-2a_t^2\Bigr)}{2(1+2u-2a_t^2-2a_0^2)\sqrt{-\frac{1}{4}-u+a_t^2+a_0^2}}t+\mathcal{O}(t^2)\right\}.
\end{equation} 

\vspace{.5cm}

 \bibliographystyle{JHEP}
 \bibliography{biblio}

\end{document}